\documentstyle[preprint,eqsecnum,aps]{revtex}
\begin{document}
\draft

\title{Gravitational field and equations of motion \\
of compact binaries to 5/2 post-Newtonian order}
\author{Luc Blanchet, Guillaume Faye and B\'en\'edicte Ponsot}
\address{D\'epartement d'Astrophysique Relativiste et de Cosmologie,\\
       Centre National de la Recherche Scientifique (UPR 176),\\
       Observatoire de Paris, 92195 Meudon Cedex, France}
\date{29 avril 1998}
\maketitle
\begin{abstract}
We derive the gravitational field and equations of motion of compact binary 
systems up to the 5/2 post-Newtonian approximation of general relativity 
(where radiation-reaction effects first appear). The approximate 
post-Newtonian gravitational field might be used in the problem of initial 
conditions for the numerical evolution of binary black-hole space-times. On 
the other hand we recover the Damour-Deruelle 2.5PN equations of motion of 
compact binary systems. Our method is based on an expression of the 
post-Newtonian metric valid for general (continuous) fluids. We substitute 
into the fluid metric the standard stress-energy tensor appropriate for a 
system of two point-like particles. We remove systematically the infinite 
self-field of each particle by means of the Hadamard partie finie 
regularization.
\end{abstract}

\section{Introduction}

The two purposes of the present paper are:

(1) To obtain the gravitational field generated by a system of two point-like 
particles up to the so-called 5/2 post-Newtonian (2.5PN) order included, i.e. 
the order $(v/c)^5$ where $v$ denotes a typical value of the orbital velocity 
of the system. The 2.5PN field may be useful for setting up initial conditions 
for the numerical study of the coalescence of two compact (neutron stars or 
black holes) objects \cite{SSSTWW97,BCTh98}. 

(2) To derive from the gravitational field the Damour-Deruelle 
\cite{BeDD81,DD81a,Dthese,D82,D83a} equations of motion of (compact) binary 
systems at the same 2.5PN order. Because the 2.5PN term in the equations of 
motion represents the dominant contribution of the radiation reaction force, 
the Damour-Deruelle equations play a crucial role in theoretically accounting 
for the decreasing of the orbital period of the binary pulsar PSR 1913+16 
\cite{TFMc79,TW82,DT91,T93}. 

In addition, the present paper is motivated by the current development of the 
future gravitational-wave observatories LIGO and VIRGO. Specifically the aim 
is to derive with sufficient post-Newtonian precision the dynamics of 
inspiralling compact binaries (which are among the most interesting sources 
to be detected by LIGO and VIRGO). Numerous authors 
\cite{3mn,CFPS93,FCh93,CF94,TNaka94,P95} have shown that the orbital phase of 
inspiralling compact binaries should be computed for applications in 
LIGO/VIRGO up to (at least) the 3PN relative order. Resolving this problem 
requires {\it in particular} the binary's equations of motion at 3PN order, 
since they permit to derive the binary's 3PN energy entering the 
left-hand-side of the energy balance equation on which rests the derivation 
of the phase. They are also needed for the computation of the 3PN 
gravitational flux entering the right-hand-side of the balance equation. Thus 
the 2.5PN equations of motion derived in \cite{BeDD81,DD81a,Dthese,D82,D83a} 
and in the present paper are not quite sufficient for the problem of 
inspiralling compact binaries, but the method we propose should permit to 
tackle in future work the problem of generalization to the next 
3PN order (see \cite{JS98} for an attempt at solving this problem).

The dynamics of a binary system of point-like particles modelling compact 
objects was investigated successfully up to 2.5PN order by Damour, Deruelle 
and collaborators \cite{BeDD81,DD81a,Dthese,D82,D83a}, using basically a 
post-Minkowskian approximation scheme (i.e. $G\to 0$). 

In a first paper by Bel {\it et al} \cite{BeDD81} (see also \cite{WeG79}), 
the gravitational field and the equations of motion are obtained in 
algebraically closed-form to the second  post-Minkowskian order ($G^2$): the 
field equations in harmonic coordinates are solved at first order by 
integrating the matter stress-energy tensor suitable to point-like sources 
(i.e. involving delta functions), and then the second-order gravitational 
field is constructed by iteration. The divergencies which arise due to the 
assumption of point-like particles are cured by means of a regularization 
process based on the Hadamard partie finie \cite{Hadam} (see \cite{Sellier} 
for an entry to mathematical literature). The equations of motion are 
obtained equivalently from the harmonicity condition to be satisfied by the 
metric, from the conservation of the (regularized) stress-energy tensor, or 
from the regularized geodesic equations.

In sequential papers \cite{DD81a,Dthese} the post-Minkowskian equations of 
motion are developed up to the order $G^2/c^5$, i.e. neglecting any term of  
the order $1/c^6$ when $c\to \infty$ {\it and} any term of the order $G^3$ 
when $G\to 0$. However, it is well known \cite{Edd23,WalW80} that 
in order to obtain the complete equations of motion to the dominant 2.5PN 
order of radiation reaction, the latter precision is not sufficient because 
of the occurence of terms coming from the third post-Minkowskian metric 
($G^3$) which contribute to both 2PN ($1/c^4$) and 2.5PN ($1/c^5$) 
approximations. These terms of orders $G^3/c^4$ and $G^3/c^5$ have been 
added by Damour \cite{D82,D83a}, thereby completing the 2.5PN binary's 
equations of motion. Let us refer to the above derivation of the dynamics of 
a binary system as the ``post-Minkowskian'' approach.

When obtaining the cubic terms within the post-Minkowskian approach 
\cite{D82,D83a} the two objects are not described by standard delta functions 
but rather by a Riesz kernel \cite{Riesz} depending on some complex parameter 
$A$. For non-zero $A$ this kernel has an infinite spatial extension, but 
reduces to the Dirac distribution when $A\to 0$. The metric is defined by 
complex analytic continuation from the $A$-dependent post-Minkowskian 
iteration. The physical equations of motion corresponding to point-like 
particles are obtained from the $A$-dependent metric by taking the limit 
$A\to 0$ at the end of the computation. It has been proved \cite{D82,D83a} 
that the limit exists up to the 2.5PN approximation (no poles $\sim 1/A$ 
develop to this order \cite{N1}).

The motion of two particles to the 2PN order (neglecting the 
radiation-reaction 2.5PN terms) is conservative, i.e. there exist ten 
integrals of motion corresponding to the Newtonian notions of energy, linear 
momentum, angular momentum and center of mass position. It has been shown 
\cite{DD81b,D83a} that the constants of motion can be recovered by variation 
of a generalized Lagrangian depending on the positions, velocities and 
accelerations of the bodies (recall that generically, i.e. in most coordinate 
systems, a 2PN Lagrangian depends upon accelerations \cite{DS85}). Adding up 
the radiation-reaction terms, one finds that the previously obtained binary's 
2PN energy decreases with time, and that there is quantitative agreement with 
the standard quadrupole formula \cite{D83b,D83a} and with the observations of 
the binary pulsar \cite{TFMc79,TW82,DT91,T93}. 

Moreover, there have been two other lines of work which led to the complete 
2.5PN dynamics of binary systems. One of these alternative approaches is 
based on  the canonical Hamiltonian formulation of general relativity and the 
model of point-like sources. Such ``Hamiltonian'' approach was developed at 
the 2PN level  in early works \cite{O74a,O74b} (see also \cite{O73,OKH75}) but 
completely understood only later \cite{DS85} (see \cite{DGKS88} for a review). 
Sch\"afer \cite{S85,S86} completed the Hamiltonian approach to include the 
2.5PN radiation-reaction terms (more recently the 3.5PN radiation-reaction 
terms have also been worked out \cite{JS97}). The other method gives up the 
model of point-like sources and assumes from the start that the two bodies are 
extended, spherically symmetric and made of perfect fluid. Within such an 
``extended-body'' approach the 2.5PN equations of motion were found 
\cite{Kop85,GKop86} to be the same as obtained within the other treatments 
dealing with point-like particles (in particular the equations depend only on 
the two masses of the bodies, but not on their internal structure nor 
compactness).

In the present paper, we add what constitutes essentially a fourth approach 
to the derivation of the 2.5PN motion of binary systems, which can be 
qualified as ``post-Newtonian'', in contrast with the post-Minkowskian, 
Hamiltonian and extended-body approaches. With respect to the 
post-Minkowskian approach \cite{BeDD81,DD81a,Dthese,D82,D83a} we have 
essentially two differences:

(1) Instead of implementing a post-Minkowskian algorithm to the third order 
and performing afterwards the post-Newtonian re-expansion, we start directly 
from a post-Newtonian metric developed to 2.5PN order and which is valid for 
any continuous matter stress-energy distribution (``fluid''). Note however 
that our post-Newtonian metric is defined in terms of {\it retarded} 
(Minkowskian) potentials, and most importantly matches a far-zone metric 
satisfying the correct boundary conditions at infinity, in particular the 
no-incoming radiation condition \cite{B95}.  

(2) Instead of assuming a fictituous stress-energy tensor defined by means of 
analytic continuation using the Riesz kernel and letting the 
analytic-continuation factor going to zero at the end of the computation 
\cite{D82,D83a}, we substitute directly into the ``fluid'' metric the 
stress-energy tensor of point-like particles. This entails divergencies which 
are cured systematically by means of the Hadamard regularization 
\cite{Hadam,Sellier} (in this respect we follow \cite{BeDD81}). 

By implementing the post-Newtonian approach, we recover the Damour-Deruelle 
equations of motion \cite{DD81a,Dthese,D82,D83a}. To the investigated order 
we find that the post-Newtonian approach is well-defined and rather 
systematic.   

About the gravitational field (metric) generated by the two particles, we 
obtain an algebraically closed-form valid everywhere in space-time up to the 
2.5PN order \cite{N2}. Indeed, the most difficult term present in the metric 
at this order is a cubically non-linear term which can be explicitly 
evaluated \cite{Dunpublished,BDI95}. This yields the other motivation of the 
present paper, namely to provide the metric coefficients at 2.5PN order 
(in harmonic coordinates) in the form of some explicit, fully reduced 
functionals of the positions and (coordinate) velocities of the two masses. 
Let us point out that, very likely, the possibility of writing such a 
closed-form expression of the metric breaks down at the next 3PN order, where 
there remain some Poisson-type integrals which probably cannot be expressed 
in terms of simple functions.

The plan of the paper is as follows. We start in Section II with the 
expression (derived in Appendix A) of the 2.5PN metric valid for general 
fluid systems. In Section III we explain our method for applying the fluid 
metric to the case of point-like particles. The metric potentials involve 
three types of terms which are evaluated respectively in Sections IV, V 
and VI (the most difficult, cubic, term being obtained in Section VI). 
The results for the potentials are relegated to Appendix B. In Section VII 
we present our expression for the binary's gravitational field. In 
Section VIII we finally obtain the (Damour-Deruelle) binary's equations of 
motion.

\section{The 2.5PN metric for general fluid systems}

At the basis of our investigation is the expression of the metric generated 
by an arbitrary matter distribution described by the stress-energy tensor 
$T^{\mu\nu}$ \cite{N3}. We assume that $T^{\mu\nu}$ has a spatially compact 
support, and physically corresponds to a slowly-moving, weakly-stressed and 
self-gravitating system, in the sense respectively that 
$|T^{0i}/T^{00}|\sim \varepsilon$, $|T^{ij}/T^{00}|\sim \varepsilon^2$ and 
$U/c^2 \sim \varepsilon^2$, where $U$ denotes the Newtonian potential and 
$\varepsilon$ represents a small post-Newtonian parameter going to zero when 
the speed of light tends to infinity ($\varepsilon \sim 1/c$). Throughout 
this paper we  denote a post-Newtonian term of order $\varepsilon^n$ by 
means of the short-hand $O(n)$. 

Following \cite{BD89} it is convenient to define a mass density $\sigma$ 
which agrees in the case of stationary systems with the Tolman mass density 
to 1PN order. As it turns out, introducing such a mass density 
(and in addition the associated retarded potential) permits to formulate the 
2.5PN metric in a rather simple fashion. Defining also some current and 
stress densities we pose
\begin{mathletters}
\label{2.1}
\begin{eqnarray}
 \sigma &\equiv& {T^{00}+T^{ii}\over c^2} \ , \\
 \sigma_i &\equiv& {T^{0i}\over c} \ , \\
 \sigma_{ij} &\equiv& T^{ij} \ .
\end{eqnarray}
\end{mathletters}
The covariant conservation of the matter stress-energy tensor 
($\nabla_\nu T^{\mu\nu}=0$) entails the equations of motion and continuity, 
which read (with relative 1PN precision in the equation of continuity but 
only Newtonian precision in the equation of motion)
\begin{mathletters}
\label{2.2}
\begin{eqnarray}
\partial_t\sigma+\partial_i\sigma_i &=& {1\over c^2}\left(\partial_t
\sigma_{ii}-\sigma\partial_tU\right)+O(4) \ , \label{2.2a}  \\
\partial_t\sigma_i+\partial_j\sigma_{ij} &=& \sigma\partial_iU+O(2) \ , 
\label{2.2b}
\end{eqnarray} 
\end{mathletters}
where $U$ is given by the standard Poisson integral: $U= \Delta^{-1}\{-4\pi 
G\sigma \}$. 

Actually, it is advantageous to use rather than the instantaneous potential 
$U$, a corresponding {\it retarded} potential $V$ given by the retarded 
integral of the same source $\sigma$:

\begin{equation}
V({\bf x},t) = \Box^{-1}_R\left\{-4\pi G\sigma \right\} \equiv G
\int {d^3{\bf z}\over |{\bf x}-{\bf z}|} 
\sigma( {\bf z}, t - |{\bf x} -{\bf z}|/c ) \ . \label{2.3} 
\end{equation}
To Newtonian order we have $V=U+O(2)$ \cite{Note}. Similarly let us introduce 
the following other retarded potentials \cite{N4}:

\begin{mathletters}
\label{2.4}
\begin{eqnarray}
V_i &=& \Box^{-1}_R\left\{-4\pi G \sigma_i\right\} \ ,  \label{2.4a}
\\ 
{\hat W}_{ij} &=& \Box^{-1}_R\left\{-4 \pi G (\sigma_{ij} - \delta_{ij} 
\sigma_{kk}) - \partial_i V \partial_j V\right\} \ , \label{2.4b} \\ 
{\hat R}_i &=& \Box^{-1}_R\left\{ - 4\pi G (V\sigma_i - V_i \sigma) - 2 
\partial_k V
\partial_i V_k - {3\over 2} \partial_t V \partial_i V \right\} \ , 
\label{2.4c}\\
{\hat X} &=& \Box^{-1}_R\biggl\{ -4\pi G V \sigma_{ii} + 2 V_i \partial_t 
\partial_i V +V \partial_t^2 V  \nonumber \\ 
&&\quad\quad+{3\over 2} (\partial_t V)^2 - 2 \partial_i V_j \partial_j V_i + 
\hat{W}_{ij} \partial^2_{ij} V \biggr\} \ . \label{2.4d}
\end{eqnarray}
In addition, we shall often consider the trace of the potential 
${\hat W}_{ij}$, i.e.

\begin{equation}
{\hat W}_{ii} = \Box^{-1}_R\left\{8\pi G \sigma_{ii}-\partial_i V \partial_i 
V \right\} \ . \label{2.4e}
\end{equation}
\end{mathletters}

We are now able to express the usual covariant metric $g_{\mu\nu}$ in terms 
of these retarded potentials to order 2.5PN, by which we mean neglecting all 
the terms of order $O(8)$ in $g_{00}$, $O(7)$ in $g_{0i}$ and $O(6)$ in 
$g_{ij}$. We impose the harmonic or De Donder coordinate conditions, i.e. 
$\partial_\nu [\sqrt{-g} g^{\mu\nu}]=0$, where $g$ and $g^{\mu\nu}$ are the 
determinant and the inverse of the matrix $g_{\mu\nu}$. Actually, since we 
are working with an approximate post-Newtonian metric, the harmonic 
conditions need only to be satisfied approximately. To the 2.5PN order we 
have $\partial_\nu [\sqrt{-g} g^{0\nu}]=O(7)$ and 
$\partial_\nu [\sqrt{-g} g^{i\nu}]=O(6)$.

With these definitions the 2.5PN metric in harmonic coordinates takes the form:
\begin{mathletters}
\label{2.5}
\begin{eqnarray}
g_{00} &=& -1 + {2\over c^2}V - {2\over c^4}V^2 + {8\over c^6} \left[\hat{X}
+ V_iV_i + {V^3\over 6}\right] +O(8) \ , \label{2.5a}\\ 
g_{0i} &=& -{4\over c^3} V_i - {8\over c^5} \hat{R}_i + O(7) \ , \label{2.5b}\\
g_{ij} &=& \delta_{ij} \left( 1 + {2\over c^2} V + {2\over c^4} V^2 
\right) + {4\over c^4} \hat{W}_{ij} + O(6) \ . \label{2.5c}
\end{eqnarray}
\end{mathletters}
For the sake of completeness, this rather simple result is proved in 
Appendix A. The simplicity in the formulation is due to our introduction of 
the mass density $\sigma$ as well as the use of retarded potentials 
\cite{BD89,B95}. 

In the form (\ref{2.5}) the metric contains only ``even'' terms explicitly 
(using the standard post-Newtonian terminology), that are terms with even 
powers of $1/c$ in $g_{00}$ and $g_{ij}$, and odd powers in $g_{0i}$. Indeed 
the ``odd'' terms, which are responsible for radiation reaction forces, are 
all hidden into the definitions of the retarded potentials 
(\ref{2.3})-(\ref{2.4}), and can be made explicit by expanding the retarded 
arguments with Taylor's formula. It is important in this respect to recall 
from \cite{B95} that (\ref{2.5}) comes from the post-Newtonian expansion 
(valid only in the near zone) of some radiative metric defined globally in 
space-time and satisfying the no-incoming radiation condition at past null 
infinity. Hence the ``odd'' terms in the post-Newtonian metric (\ref{2.5}) 
correspond physically to the radiation reaction forces acting on an isolated 
system (with no source located at infinity).

At 2.5PN order the harmonic-coordinate conditions are equivalent to the 
following differential identities:
\begin{mathletters}
\label{2.6}
\begin{eqnarray}
& & \partial_t\left\{V+{1\over c^2}\left[{1\over 2}\hat{W}_{ii}+
2V^2\right]\right\} +\partial_i\left\{V_i+{2\over c^2}
\left[\hat{R}_i+VV_i\right]\right\}= O(4) \ , \label{2.6a}\\ 
& & \partial_t V_i +\partial_j\left\{{\hat W}_{ij}-{1\over 2}\delta_{ij} 
{\hat W}_{kk} \right\}=O(2) \ . \label{2.6b}
\end{eqnarray}
\end{mathletters}
These relations are in turn equivalent to the 1PN continuity equation and 
Newtonian equation of motion given by (\ref{2.2}).

The potentials $V$ and $V_i$ are generated by the compact-supported source 
densities $\sigma$ and $\sigma_i$. Similarly ${\hat W}_{ij}$ and ${\hat R}_i$ 
involve a part generated by a compact-supported source, but also a part whose 
source is a sum of quadratic products of potentials $V$ or $V_i$ and their 
space-time derivatives. We shall refer to the former part 
of ${\hat W}_{ij}$ and ${\hat R}_i$ as the compact (``C'') part, and to the 
latter as the ``$\partial V\partial V$'' or, sometimes, ``quadratic" part. As 
for ${\hat X}$, it consists of C and $\partial V\partial V$ parts like for 
${\hat W}_{ij}$ and ${\hat R}_i$, but it also contains a term of different 
structure, generated by the product of ${\hat W}_{ij}$ and  
$\partial^2_{ij}V$ [last term in (\ref{2.4d})]. This term itself can be split 
into two contributions arising respectively from the C and $\partial V
\partial V$ parts of ${\hat W}_{ij}$. Since the C part of ${\hat W}_{ij}$ is 
a compact-supported potential similar to $V$ and $V_i$, the corresponding 
term in ${\hat X}$ has actually the same structure as a 
$\partial V\partial V$ potential. On the other hand, the 
$\partial V\partial V$ part of ${\hat W}_{ij}$ generates an intrinsically 
more complicated term  in ${\hat X}$ we shall refer to as the non-compact 
(``NC'') or ``cubic'' part of the potential ${\hat X}$. Precisely, our 
definitions are
\begin{mathletters}
\label{2.7}
\begin{eqnarray}
\hat{W}_{ij} &=& \hat{W}_{ij}^{({\rm C})} + \hat{W}_{ij}^{(\partial V
\partial V)} \ , \label{2.7a}\\
\hat{R}_i &=& \hat{R}_i^{({\rm C})} + \hat{R}_i^{(\partial V\partial V)} \ , 
\label{2.7b}\\
\hat{X} &=& \hat{X}^{({\rm C})} + \hat{X}^{(\partial V\partial V)} + 
\hat{X}^{({\rm NC})} \ . \label{2.7c}
\end{eqnarray} 
\end{mathletters}
The compact parts are linear and quadratic functionals of the matter 
variables (\ref{2.1}). They read as 
\begin{mathletters}
\label{2.8}
\begin{eqnarray}
\hat{W}_{ij}^{({\rm C})} &=& \Box^{-1}_R\left\{-4 \pi G (\sigma_{ij} - 
\delta_{ij} \sigma_{kk})\right\} \ , \label{2.8a} \\
\hat{R}_i^{({\rm C})} &=& \Box^{-1}_R\left\{ - 4\pi G (V\sigma_i - 
V_i \sigma)\right\} \ , \label{2.8b} \\
\hat{X}^{({\rm C})} &=& \Box^{-1}_R\left\{ -4\pi G V \sigma_{ii}\right\} \ . 
\label{2.8c}
\end{eqnarray}
\end{mathletters}
The $\partial V\partial V$ or ``quadratic'' parts involve both quadratic and 
cubic contributions. They are
\begin{mathletters}
\label{2.9}
\begin{eqnarray}
\hat{W}_{ij}^{(\partial V\partial V)} &=& \Box^{-1}_R\left\{- \partial_i V 
\partial_j V\right\} \ , \label{2.9a} \\
\hat{R}_i^{(\partial V\partial V)} &=& \Box^{-1}_R\left\{ - 2 
\partial_k V \partial_i V_k - {3\over 2} \partial_t V \partial_i V \right\} 
\ , \label{2.9b} \\
\hat{X}^{(\partial V\partial V)} &=& \Box^{-1}_R\left\{ 2 V_i \partial_t 
\partial_i V +V \partial_t^2 V  +{3\over 2} 
(\partial_t V)^2 \right.\nonumber \\
&&\quad\quad -\left.2 \partial_i V_j \partial_j V_i + \hat{W}_{ij}^{({\rm C})} 
\partial^2_{ij} V \right\} \ . \label{2.9c}
\end{eqnarray}
\end{mathletters}
Finally the only Non-Compact part is a cubic functional given by

\begin{equation}
\hat{X}^{({\rm NC})} = \Box^{-1}_R\left\{\hat{W}_{ij}^{(\partial V\partial V)}
\partial^2_{ij} V \right\} \ . \label{2.10}
\end{equation}
In practice the latter term is the most delicate to evaluate. Our terminology 
is slightly improper, as the so-called $\partial V\partial V$ or quadratic 
potentials are as well as the NC potential generated by non-compact supported 
sources, and involve some contributions which are actually cubic in the 
matter variables. 

\section{Application to point-like particles}

To apply the general 2.5PN metric presented in the previous section 
to the case of a point-mass binary we use the matter stress-energy tensor:

\begin{equation}
T^{\mu\nu} ({\bf x},t) = \mu_1(t) v_1^\mu(t) v_1^\nu(t) 
\delta({\bf x}-{\bf y}_1(t)) + 1\leftrightarrow 2 \ . \label{3.1}
\end{equation}
In our notation the symbol $1\leftrightarrow 2$ means the same term but with 
the labels 1 and 2 exchanged; $\delta$ denotes the three-dimensional Dirac 
distribution; the trajectories of the two masses (in harmonic coordinates) 
are denoted by ${\bf y}_1(t)$ and ${\bf y}_2(t)$; the two coordinate 
velocities are ${\bf v}_1(t)=d{\bf y}_1(t)/dt$, 
${\bf v}_2(t)=d{\bf y}_2(t)/dt$, and $v_1^\mu \equiv (c,{\bf v}_1)$, 
$v_2^\mu \equiv (c,{\bf v}_2)$; $\mu_1$ represents an effective 
time-dependent mass of body 1 defined by

\begin{equation}
\mu_1(t) =\left( {m_1\over \sqrt{gg_{\rho\sigma} {\displaystyle 
\frac{v_1^\rho v_1^\sigma}{c^2}}}} \right)_1 \ , \label{3.2}
\end{equation}
$m_1$ being the (constant) Schwarzschild mass, with $g_{\rho\sigma}$ the 
 metric and $g$ its determinant \cite{N7}. 
Another useful notation is

\begin{equation}
\tilde{\mu}_1(t) = \mu_1(t)\left[1+{v_1^2 \over c^2}\right] \ , \label{3.3}
\end{equation}
where $v_1^2={\bf v}_1^2$. Both $\mu_1$ and $\tilde{\mu}_1$ reduce to the 
Schwarszchild mass at Newtonian order: $\mu_1=m_1+O(2)$ and 
$\tilde{\mu}_1=m_1+O(2)$. Then the mass, current and stress densities 
(\ref{2.1}) for two particles read

\begin{mathletters}
\label{3.4}
\begin{eqnarray}
\sigma &=& \tilde{\mu}_1 \delta ({\bf x}-{\bf y}_1) + 1\leftrightarrow 2 \ , 
\label{3.4a} \\
\sigma_i &=& \mu_1 v_1^i \delta ({\bf x}-{\bf y}_1)+ 1\leftrightarrow 2 \ , 
\label{3.4b} \\
\sigma_{ij} &=& \mu_1 v_1^i v_1^j \delta ({\bf x}
-{\bf y}_1)+ 1\leftrightarrow 2 \ . \label{3.4c} 
\end{eqnarray}
\end{mathletters}

The stress-energy tensor of point-masses depends on the values of the metric 
coefficients at the very location of the particles. However the metric 
coefficients there become infinite and, consequently, we must supplement the 
model of stress-energy tensor (\ref{3.1}) by a prescription for giving a 
sense to the notion of the field sitting on the particle. In other words, we 
need a regularization procedure in order to remove the infinite self-field of 
point-like sources. The choice of one or another regularization procedure 
represents ({\it a priori}) an integral part of the choice of physical model 
for describing the particles. 
In the present paper we shall employ the Hadamard regularization 
\cite{Hadam,Sellier} based on the partie finie of functions admitting a 
special (``tempered'') type of singularity. For discussion and justification 
of the use of the Hadamard regularization in the context of equations of 
motion in general relativity see \cite{BeDD81,Dthese,S85,BDI95,JS98,Jara97}.

Let us consider the class of functions $F$ depending on the field point 
${\bf x}$ as well as on two source points ${\bf y}_1$ and ${\bf y}_2$, and 
admitting when the field point approaches one of the source points 
($r_1 = |{\bf x}-{\bf y}_1| \to 0$ for instance) an expansion of the type

\begin{equation}
F({\bf x};{\bf y}_1,{\bf y}_2)=\sum_{\scriptstyle -k_0\leq k\leq 0} 
r_1^k f_k({\bf n}_1;{\bf y}_1,{\bf y}_2)+O(r_1)  \label{3.5}
\end{equation}
(where $k \in Z\!\!\!Z$). We define the value of the function $F$ at the 
source point 1 (and similarly at the source point 2) to be the so-called 
Hadamard partie finie, which is the average, with respect to the direction 
${\bf n}_1=({\bf x}-{\bf y}_1)/r_1$ of approach to point 1, of the term with 
zeroth power of $r_1$ in (3.5). Namely \cite{N5},

\begin{equation}
(F)_1 \equiv F({\bf y}_1;{\bf y}_1,{\bf y}_2)\equiv\int {d\Omega({\bf n}_1) 
\over 4\pi} f_0({\bf n}_1;{\bf y}_1,{\bf y}_2)\ . \label{3.6}
\end{equation}
Furthermore, we use the Hadamard partie finie to give a sense to the spatial 
integral of the product of F and the Dirac delta function at point 1 (since 
$F$ is singular on the support of the Dirac function). Indeed, we define 

\begin{equation}
\int d^3{\bf x} F({\bf x};{\bf y}_1,{\bf y}_2) \delta({\bf x}-{\bf y}_1)  
\equiv (F)_1 \ , \label{3.7}
\end{equation}
where $(F)_1$ is given by (\ref{3.6}). 

As a (trivial) example of the use of the Hadamard regularization, consider 
the potentials $V$ and $V_i$ to Newtonian order, given by

\begin{mathletters}
\label{3.8}
\begin{eqnarray}
V &=& {Gm_1\over r_1}+O(2)+1\leftrightarrow 2 \ , \label{3.8a}\\
V_i &=& {Gm_1\over r_1}v_1^i+O(2)+1\leftrightarrow 2 \ . \label{3.8b}
\end{eqnarray}
\end{mathletters}
They are infinite at point 1, but after applying the rule (\ref{3.6}) we find

\begin{mathletters}
\label{3.9}
\begin{eqnarray}
(V)_1 &=& {Gm_2\over r_{12}}+O(2) \ , \label{3.9a}\\
(V_i)_1 &=& {Gm_2\over r_{12}}v_2^i+O(2) \ ,  \label{3.9b}
\end{eqnarray}
\end{mathletters}
where $r_{12}=|{\bf y}_1-{\bf y}_2|$ is the distance between the particles 
\cite{N6}. Of course $(V)_1$ agrees with the standard Newtonian result. 
Applying the rule (\ref{3.7}) we have, for instance,  

\begin{equation}
{1\over 2}\int d^3{\bf x}~\sigma V = {Gm_1m_2\over r_{12}}+O(2) \ ,  
\label{3.10}
\end{equation}
also in agreement with the Newtonian result.

We shall derive the binary's equations of motion in the so-called 
order-reduced form, by which we mean that in the final result all 
accelerations (and time-derivatives of accelerations) are replaced 
consistently with the approximation by the explicit functionals of the 
positions and velocities  as given by the (lower-order) equations. So, in 
order to derive the 2.5PN equations of motion (and also the metric), we use 
the less accurate 1.5PN equations, given in harmonic coordinates by

\begin{eqnarray}
{dv_1^i\over dt} =&-&{Gm_2\over r_{12}^2} n_{12}^i \left\{1+
{1\over c^2}\left[ - 5{Gm_1\over r_{12}} - 4{Gm_2\over r_{12}} +
v^2_1 + 2v^2_2 - 4(v_1v_2) -{3\over 2} (n_{12}v_2)^2 \right]\right\}\nonumber\\
 &+&{Gm_2\over c^2r_{12}^2} v^i_{12} \left[4(n_{12}v_1) -3(n_{12}v_2)\right] 
+O(4) \ , \label{3.11}
\end{eqnarray}
with ${\bf n}_{12}=|{\bf y}_1-{\bf y}_2|/r_{12}$ and ${\bf v}_{12}={\bf v}_1-
{\bf v}_2$; scalar products are denoted with parenthesis, e.g. $(n_{12}v_1)=
{\bf n}_{12}.{\bf v}_1$. The acceleration of body 2 
is obtained by exchanging the labels 1 and 2 (remembering that ${\bf n}_{12}$ 
and ${\bf v}_{12}$ change sign in this operation).

\section{The compact parts of potentials}

In this section we derive the compact-supported potentials $V$ and $V_i$, and 
the compact-supported parts of the other potentials, 
${\hat W}_{ij}^{({\rm C})}$, ${\hat R}_i^{({\rm C})}$ and 
${\hat X}^{({\rm C})}$ defined by (\ref{2.8}), for a binary system described 
by the stress-energy tensor (\ref{3.1}) and the regularization (3.6).  
We need $V$ to relative 2.5PN order, $V_i$ to 1.5PN order, and the other 
compact potentials to 0.5PN order only. The main task is the computation of 
$V$, to which we focus mainly our attention. 
By Taylor-expanding at 2.5PN order the retardation inside the integral 
(\ref{2.3}) and using the mass density $\sigma$ in the form (\ref{3.4a}), we 
get 

\begin{eqnarray}
V = G \left\{ {{\tilde \mu}_1\over r_1} \right.&-&{1\over c}\partial_t 
({\tilde \mu}_1)
+{1\over 2c^2} \partial_t^2 ({\tilde \mu}_1 r_1) 
- {1\over 6c^3} \partial_t^3 ({\tilde \mu}_1 r_1^2)\nonumber\\
&+& \left. {1\over 24c^4} \partial_t^4 ({\tilde \mu}_1 r_1^3) 
- {1\over 120c^5} \partial_t^5 ({\tilde \mu}_1 r_1^4) \right\}
+ O(6) + 1 \leftrightarrow 2  \ . \label{4.1}
\end{eqnarray}
We recall that the effective mass ${\tilde \mu}_1$ given by 
(\ref{3.2})-(\ref{3.3}) is a function of time only.  

We start by deriving  ${\tilde \mu}_1$ to 2.5PN order. Inserting the metric 
coefficients (\ref{2.5}) into the expressions (\ref{3.2})-(\ref{3.3}), we 
obtain 

\begin{eqnarray}
\tilde{\mu}_1 &=& m_1 \left\{ 1 + {1\over c^2} \left [-(V)_1 + {3\over 2}
v_1^2\right] \right.\nonumber\\
&+& \left.{1\over c^4} \left[-2({\hat W}_{ii})_1 + {1\over 2} (V^2)_1 + 
{1\over 2} (V)_1 v_1^2
- 4 (V_i)_1 v_1^i + {7\over 8} v_1^4\right] \right\} +O(6) \ , \label{4.2} 
\end{eqnarray}
where all the potentials are to be evaluated at the location of body 1, using 
the rule (\ref{3.6}). We proceed iteratively. The first step consists in 
inserting into (\ref{4.2}) the potential $V$ at body 1 to Newtonian order 
(or, rather, 0.5PN order), which is simply the Newtonian result (\ref{3.9a}). 
This yields ${\tilde \mu}_1$ to 1.5PN order:

\begin{equation}
\tilde{\mu}_1 = m_1\left\{1 + {1\over c^2} \left [-{Gm_2\over r_{12}}
+ {3\over 2} v_1^2 \right] \right\} + O(4) \ . \label{4.3}
\end{equation}
Since the time-derivative of ${\tilde \mu}_1$ starts at 1PN order, namely

\begin{equation}
{\dot{\tilde \mu}}_1 = {Gm_1m_2\over c^2r_{12}^2}\left[-2(n_{12}v_1)
-(n_{12}v_2)\right] + O(4) \ , \label{4.4}
\end{equation}
we see that the first odd power of $1/c$ in $V$ arises at 1.5PN order. 
Furthermore, using the constancy of the center of mass velocity, one can 
check that the first odd term in the gradient of $V$ arises at 2.5PN order 
(it contributes to the dominant radiation reaction force). From (\ref{4.1}) 
and (\ref{4.3}) we deduce the value of $(V)_1$ up to 1.5PN order:

\begin{eqnarray}
(V)_1 &=& {Gm_2\over r_{12}}\left\{1+{1\over c^2}\left[-{3Gm_1\over 2 r_{12}}+
2v_2^2-{1\over 2}(n_{12}v_2)^2\right]\right\}\nonumber\\
&+&{4 G^2m_1m_2\over 3 c^3r_{12}^2} (n_{12}v_{12}) + O(4) \ . \label{4.5}
\end{eqnarray}
In addition to $(V)_1$, we need $(V_i)_1$ already given by (\ref{3.9b}), and 
the value at point 1 of the trace of ${\hat W}_{ij}$ to 0.5PN order. The 
trace ${\hat W}_{ii}$ is much simpler than the potential itself, and from 
(\ref{2.4e}) we derive the expression

\begin{eqnarray}
\hat{W}_{ii} &=& \Delta^{-1}\left\{8\pi G \left(\sigma_{ii} -
{1\over 2}\sigma V\right)\right\} -{1\over 2}V^2\nonumber\\
&+&{2G\over c}{d\over dt}\int d^3{\bf x}\left(\sigma_{ii} -
{1\over 2}\sigma V\right)+O(2) \ . \label{4.6}
\end{eqnarray} 
Under this form all integrals are compact-supported; at this order, we can 
insert $V=U+O(2)$. The odd term $O(1)$ is a mere function of time. From 
(\ref{4.6}) we get immediately 

\begin{equation}
({\hat W}_{ii})_1 = {Gm_2\over r_{12}}\left[{Gm_1\over r_{12}}-
{Gm_2\over 2r_{12}}-2v_2^2\right]
-{2G^2m_1m_2\over cr_{12}^2}(n_{12}v_{12})+ O(2) \ .  \label{4.7}
\end{equation}
The effective mass ${\tilde \mu}_1$ at 2.5PN order is readily obtained from 
the previous relations: 

\begin{eqnarray}
\tilde{\mu}_1 &=& m_1\biggl\{1 + {1\over c^2} \left [-{Gm_2\over r_{12}}
+ {3\over 2} v_1^2 \right]   \nonumber\\ 
&&\quad +{1\over c^4} \left[ {Gm_2\over r_{12}} \left({1\over 2}
v_1^2 - 4(v_1v_2)
+ 2 v_2^2 + {1\over 2}(n_{12}v_2)^2 - {1\over 2}{Gm_1\over r_{12}}+
{3\over 2}{Gm_2\over r_{12}} \right) + {7\over 8} v_1^4 \right]\nonumber\\
&&\quad +{8 G^2m_1m_2\over 3 c^5r_{12}^2}(n_{12}v_{12}) \biggr\} +O(6) \ , 
\label{4.8}
\end{eqnarray}
from which we straightforwardly deduce $V$ to 2.5PN order. The only point is 
to compute the numerous (up to five) time-derivatives of $r_1$. This gives 
rise to many terms depending on the acceleration and its time-derivatives, 
that we reduce order by order by means of the binary's 1.5PN equations of 
motion (since an acceleration already arises in the 1PN term of $V$). Once 
fully reduced, the result for $V$ is still lengthy so we relegate it 
(together with all the relevant results for the potentials) in Appendix B.

The potential $V_i$ to 1.5PN order, and the other compact potentials 
${\hat W}_{ij}^{({\rm C})}$, ${\hat R}_i^{({\rm C})}$ and 
${\hat X}^{({\rm C})}$ to 0.5PN order, are obtained in the same way. As an 
example we give

\begin{equation}
{\hat W}_{ij}^{({\rm C})} = {Gm_1\over r_1}\left(v_1^iv_1^j-
\delta^{ij}v_1^2\right)
+{G^2m_1m_2\over cr_{12}^2}\left[n_{12}^{(i}v_{12}^{j)}-
\delta^{ij}(n_{12}v_{12})\right]
+O(2)+1\leftrightarrow 2 \ . \label{4.9}
\end{equation}

\section{The quadratic parts of potentials}

By the definition (\ref{2.9}) the quadratic or $\partial V\partial V$ 
potentials have their sources made of quadratic products of (derivatives of) 
the compact-supported potentials $V$, $V_i$ and ${\hat W}_{ij}^{({\rm C})}$. 
All the $\partial V\partial V$ potentials are to be computed to 0.5PN order, 
which means in particular that we can replace in the sources $V$ and $V_i$ 
by the Newtonian-like potentials $U$ and $U_i$ (but we must beware of the 
fact that ${\hat W}_{ij}^{({\rm C})}$ given by (\ref{4.9}) involves a $1/c$ 
correction). For all the $\partial V\partial V$ potentials we proceed 
similarly. We work out the sources using (\ref{3.8}) and (\ref{4.9}) and 
obtain some ``self'' terms, proportional to $m_1^2$ and $m_2^2$, together 
with some ``interaction'' terms, proportional to $m_1m_2$. Time-derivatives 
are changed to spatial derivatives thanks to  
$\partial_t(1/r_1)=v_1^i\partial_{1i}(1/r_1)$ and 
$\partial_t^2(1/r_1)=a_1^i\partial_{1i}(1/r_1)+
v_1^iv_1^j\partial_{1ij}(1/r_1)$, $a_1^i$ denoting the acceleration and 
$\partial_{1i}$ the partial derivative with respect to $y_1^i$. In the 
interaction terms, we leave the spatial derivatives un-expanded, whereas, in 
the self terms, they are developed and ``factorized'' out in front of the 
terms. In the latter operation, we should remember that within the standard 
distribution theory the second spatial derivative of $1/r_1$ involves a 
distributional 
term \cite{EK85,Sellier,Jara97}: 

\begin{mathletters}
\label{5.1}
\begin{eqnarray}
\Delta \left({1\over r_1}\right) &=& -4\pi \delta ({\bf x}-{\bf y}_1) \ , 
\label{5.1a}\\
\partial^2_{ij}\left({1\over r_1}\right) &=& {3n_1^in_1^j-
\delta^{ij} \over r_1^3}-
{4\pi\over 3}\delta^{ij} \delta ({\bf x}-{\bf y}_1) \ . \label{5.1b}
\end{eqnarray}
\end{mathletters}
Two examples of such a treatment of sources are

\begin{mathletters}
\label{5.2}
\begin{eqnarray}
\partial_i V\partial_j V &=& {G^2m_1^2\over 8}\left(\partial^2_{1ij}+
\delta^{ij}\Delta_1\right)\left({1\over 
r_1^2}\right)\nonumber\\ 
&+& G^2m_1m_2 \partial_{1i} \partial_{2j}\left({1\over r_1r_2}\right)+O(2)+
1\leftrightarrow 2  \ , \label{5.2a}\\
V\partial_t^2 V &=& G^2m_1^2\left[{1\over 8}\left(4a_1^i\partial_{1i}+
3v_1^{ij}\partial^2_{1ij}-v_1^2 
\Delta_1 \right) \left({1\over r_1^2}\right)
-{4\pi\over 3}{v_1^2\over r_1}\delta({\bf x}-{\bf y}_1)\right]\nonumber\\ 
&+&G^2m_1m_2\left(a_1^i\partial_{1i}+
v_1^{ij}\partial^2_{1ij}\right)\left({1\over r_1r_2}\right)+O(2)+
1\leftrightarrow 2 \ . \label{5.2b}
\end{eqnarray}
\end{mathletters}

We apply the Poisson integral on the source terms treated in the previous 
manner. Consider firstly the distributional terms, such as the one in the 
self part of $V\partial_t^2 V$. Although this term is ill-defined, because 
involving the product of the Dirac distribution $\delta({\bf x}-{\bf y}_1)$ 
by $1/r_1$ which is singular when ${\bf x}\rightarrow {\bf y}_1$, the Poisson 
integral is computed unambiguously with the help of the Hadamard 
regularization (\ref{3.7}), yielding zero in this case:

\begin{eqnarray}
\Delta^{-1}\left[-{4\pi\over r_1}\delta({\bf x}-{\bf y}_1)\right]&=&
\int {d^3{\bf z}\over |{\bf x}-{\bf z}||{\bf z}-{\bf y}_1|}
\delta({\bf z}-{\bf y}_1)\nonumber\\
&=&\left({1\over |{\bf x}-{\bf z}|
|{\bf z}-{\bf y}_1|}\right)_{{\bf z}={\bf y}_1}=0  \ . \label{5.3}
\end{eqnarray}
For the computation of all non-distributional terms in the $\partial V
\partial V$ potentials, we take the example of 

\begin{eqnarray}
\hat{W}_{ij}^{(\partial V\partial V)} &=& \Box^{-1}_R\left\{-\partial_i V 
\partial_j V \right\} \nonumber \\
&=& \Delta^{-1}\left\{- \partial_i V \partial_j V\right\} +{1\over 4\pi c}
{d\over dt}\int d^3{\bf x}~\{-\partial_i V 
\partial_j V\}+O(2)  \ , \label{5.4}
\end{eqnarray}
whose ``source'' is given by (\ref{5.2a}). The Poisson integral of the 
self-terms  can be readily deduced from $\Delta (\ln r_1)=1/r_1^2$; on the 
other hand, that of the interaction terms is obtained by solving the 
elementary Poisson equation

\begin{equation}
\Delta g = {1\over r_1 r_2} \ .  \label{5.5}
\end{equation}
The solution is known \cite{N8}:
 
\begin{equation}
g = \ln S \ ; \qquad S\equiv r_1+r_2+r_{12} \ . \label{5.6}
\end{equation} 

The computation of the $1/c$-term in (\ref{5.4}) involves essentially the 
spatial integral of $1/r_1r_2$. Since it is divergent due to the bound at 
infinity (i.e. when $r\equiv |{\bf x}|\to \infty$), we first compute the 
finite integral defined by integration over a ball of constant finite radius 
${\cal R}$. By writing the integrand as $1/r_1r_2=\Delta g$ and using the 
Gauss theorem, we transform the integral into a surface integral over the 
sphere of radius ${\cal R}$,

\begin{equation}
\int_{|{\bf x}|\leq {\cal R}} {d^3{\bf x}\over r_1r_2}
=\int_{|{\bf x}|\leq {\cal R}} d^3{\bf x}~\Delta g
=\int_{r={\cal R}}d\Omega~(r^2\partial_rg) \ , \label{5.7}
\end{equation}
with $\partial_r\equiv n^i\partial_i$. Into the latter surface integral we 
can replace the function $g$ by its expansion at infinity computed from 
(\ref{5.6}): $g=\ln (2r)+[-(ny_1)-(ny_2)+r_{12}]/2r+O(1/r^2)$. Neglecting the 
terms which die out in the limit ${\cal R}\to\infty$, we get

\begin{equation}
-{1\over 4\pi}\int_{|{\bf x}|\leq {\cal R}} {d^3{\bf x}\over r_1r_2}
=-{\cal R}+\frac{r_{12}}{2}+O\left({1\over {\cal R}}\right) \ . \label{5.8}
\end{equation}
As we can see, the divergent part of the integral is simply a constant, which 
will therefore vanish after application of the spatial derivatives 
$\partial_{1i}\partial_{2j}$ in front of the term. This shows that we are 
allowed to use in this computation the finite coefficient in the 
right-hand-side of (\ref{5.8}), which is nothing but the partie finie in the 
sense of Hadamard of the initially divergent integral \cite{N9}. Setting 
${\bf y}_1={\bf y}_2$ in (\ref{5.8}), we infer that the divergent integral of 
$1/r_1^2$ can be replaced by zero. Then, (\ref{5.8}) together with the last 
fact lead to

\begin{equation}
-{1\over 4\pi}\int d^3{\bf x}~\partial_i V \partial_j V
={G^2m_1m_2\over r_{12}}\left(n_{12}^{ij}-\delta^{ij}\right)+O(2) \ . 
\label{5.9}
\end{equation}
Gathering those results, we thereby obtain the looked-for potential as 

\begin{eqnarray}
\hat{W}_{ij}^{(\partial V\partial V)} = &-&{G^2m_1^2\over 8}\left( 
\partial^2_{ij}\ln r_1+{\delta^{ij}\over 
r_1^2}\right) - G^2m_1m_2 \,_ig_j \nonumber\\
&+&{G^2m_1m_2\over cr_{12}^2}\left[n_{12}^{(i}v_{12}^{j)}-
{1\over 2}  \left(3n_{12}^{ij}- \delta^{ij}\right)(n_{12}v_{12})\right]
+O(2)+1\leftrightarrow 2 \ , \label{5.10}
\end{eqnarray}
where we have ${}_ig_j\equiv \partial_{1i}\partial_{2j}g$. The first two 
terms are in agreement with a result of \cite{BDI95}. All the $\partial V
\partial V$-potentials are calculated in this way to obtain the complete 
expressions of potentials presented in Appendix B.

Ending this section we list some formulas which are useful in the derivation 
of the $\partial V\partial V$-potentials, and even elsewhere. The first-order 
spatial derivatives of $g$ read 

\begin{mathletters}
\label{5.11}
\begin{eqnarray}
{}_ig &\equiv& \partial_{1i}g = {-n_1^i+n_{12}^i\over S} \ , \label{5.11a}\\
g_i &\equiv& \partial_{2i}g = {-n_2^i-n_{12}^i\over S} \ , \label{5.11b}\\
\partial_ig &=& -{}_ig-g_i={n_1^i+n_2^i\over S}  \ , \label{5.11c}
\end{eqnarray}
\end{mathletters}
and second-order spatial derivatives are (with $n^{ij}_{12}\equiv 
n^i_{12} n^j_{12}$)

\begin{mathletters}
\label{5.12}
\begin{eqnarray}
{}_{ij}g &\equiv& \partial^2_{1ij}g =-{n_{12}^{ij}-\delta^{ij}\over r_{12}S} -
{n_1^{ij}-\delta^{ij}\over r_1S}-{(n_{12}^i-
n_1^i)(n_{12}^j-n_1^j)\over S^2} \ , \label{5.12c}\\
g_{ij} &\equiv& \partial^2_{2ij}g =-{n_{12}^{ij}-\delta^{ij}\over r_{12}S} -
{n_2^{ij}-\delta^{ij}\over r_2S}-
{(n_{12}^i+n_2^i)(n_{12}^j+n_2^j)\over S^2} \ , \label{5.12b}\\
{}_ig_j &\equiv& \partial_{1i}\partial_{2j}g = {n_{12}^{ij}-
\delta^{ij}\over r_{12}S}+{(n_{12}^i-
n_1^i)(n_{12}^j+n_2^j)\over S^2} \ . \label{5.12a}
\end{eqnarray}
\end{mathletters}
Contracting with the Kronecker $\delta^{ij}$, the relations (\ref{5.12}) 
become ${}_{ii}g=\Delta_1g=1/r_1r_{12}$, $g_{ii}=\Delta_2g=1/r_2r_{12}$, and, 
more interestingly,  
\begin{equation}
{}_ig_i = {1\over 2}\left({1\over r_1r_2}-{1\over r_2r_{12}}-
{1\over r_1r_{12}}\right) \ . \label{5.13}
\end{equation}
This simple result is a straightforward consequence of the helpful formulas 

\begin{mathletters}
\label{5.14}
\begin{eqnarray}
{1+(n_1n_2)\over S} &=& {r_1+r_2-r_{12}\over 2r_1r_2} \ , \label{5.14a}\\
{1-(n_1n_{12})\over S} &=& {r_1+r_{12}-r_2\over 2r_1r_{12}} \ , \label{5.14b}\\
{1+(n_2n_{12})\over S} &=& {r_2+r_{12}-r_1\over 2r_2r_{12}} \ . \label{5.14c}
\end{eqnarray}
\end{mathletters}
Finally, we find the values of $g$ and its derivatives at the location of 
body 1 (say) according to the Hadamard regularization (3.6):  
$(g)_1=\ln(2r_{12})$, together with

\begin{mathletters}
\label{5.15}
\begin{eqnarray}
({}_ig)_1 &=& {n_{12}^i\over 2r_{12}} \ ; \ \qquad \quad \quad
(g_i)_1 = -{n_{12}^i\over r_{12}} \ , \label{5.15a}\\
({}_ig_j)_1 &=& {-\delta^{ij}+2n_{12}^{ij}\over 2r_{12}^2} \ ; \quad
({}_ig_i)_1 = -{1\over 2r_{12}^2} \ , \label{5.15b}\\
({}_{ij}g)_1 &=& {\delta^{ij}-3n_{12}^{ij}\over 4r_{12}^2}\ ; \quad\quad 
(g_{ij})_1 = {\delta^{ij}-2n_{12}^{ij}\over r_{12}^2} \ . \label{5.15c}
\end{eqnarray}
\end{mathletters}
These formulas are extensively used when getting the potentials at 1 
(see Appendix B).

\section{The non-compact potential}

The so-called non-compact potential is defined by (\ref{2.10}) as the 
retarded integral of a source composed of the product of a $V$-type potential 
and  $\hat{W}_{ij}^{(\partial V\partial V)}$, the latter potential being 
itself given as the retarded integral of a source made of a quadratic product 
of $V$'s. Due to this purely cubic structure, one would expect {\it a priori} 
that the computation of the non-compact term represents a non-trivial task, 
and even that it is not at all guaranteed that this term can be expressible 
with the help of simple algebraic functions (algebraically closed-form). 
Rather surprisingly, the NC potential turns out to accept an algebraically 
closed-form up to 0.5PN order. As a result, one can find its explicit 
expression, valid for any source point ${\bf x}$ (the value when the source 
point sits on a particle ${\bf y}_{1,2}$ following from the regularization 
process). To Newtonian order the closed-form expression of the NC term has 
already been obtained in \cite{Dunpublished,BDI95} by combining some 
technical results derived earlier in \cite{Car65,O73,S87}. We shall present 
here a slightly different but totally  equivalent form of 
$\hat{X}^{({\rm NC})}$ at Newtonian order, and add to this the 0.5PN 
correction. Very likely the 1PN and higher corrections in 
$\hat{X}^{({\rm NC})}$ do not admit any algebraically closed-form all over 
space-time, but the regularized values at the location of the two bodies can 
probably be carried out explicitly (these values are needed when 
investigating the equations of motion).

To 0.5PN order the non-compact potential reads as

\begin{eqnarray}
\hat{X}^{({\rm NC})} &=& \Box^{-1}_R \left\{ \hat{W}_{ij}^{(\partial V
\partial V)} \partial^2_{ij} V \right\} \nonumber\\
&=& \Delta^{-1} \left\{ \hat{W}_{ij}^{(\partial V\partial V)} \partial^2_{ij} 
V \right\} 
+{1\over 4\pi c}{d\over dt}\int d^3{\bf x} \left\{ \hat{W}_{ij}^{(\partial V
\partial V)} \partial^2_{ij} V\right\} + O(2) \ , \label{6.1}
\end{eqnarray}
where to this order $V$ can be replaced by (\ref{3.8a}) and 
$\hat{W}_{ij}^{(\partial V\partial V)}$ by (\ref{5.10}). The cubic source is 
easily obtained thanks to (\ref{3.8a}) and (\ref{5.10}). Hence we arrive at

\begin{eqnarray}
{\hat W}_{ij}^{(\partial V\partial V)}\partial^2_{ij}V &=& {G^3m_1^3\over 2} 
\left[ {1 \over r_1^5} +{\pi \over r_1^2} \delta({\bf x}-{\bf y}_1) \right] 
\nonumber\\
&+& G^3m_1^2m_2\left\{{\pi\over 2}{\delta({\bf x}-{\bf y}_2)\over r_{1}^2}-
{1\over 8}\partial^2_{ij}\left({1\over 
r_2}\right)\partial^2_{ij}\ln r_1-
2\partial^2_{ij}\left({1\over r_1}\right){}_ig_j\right\} \nonumber\\
&+&{G^3m_1^2m_2\over cr_{12}^2}\left[n_{12}^iv_{12}^j-
{1\over 2}\left(3n_{12}^{ij}-
\delta^{ij}\right)(n_{12}v_{12})\right] 
\partial^2_{1ij}\left({2\over r_1}\right)\nonumber\\
&+&O(2)+1\leftrightarrow 2 \ . \label{6.2}
\end{eqnarray}

We compute the Poisson integral of (\ref{6.2}). The (ill-defined) 
distributional term in the self part ($\propto m_1^3$) of (\ref{6.2}) is 
treated unambiguously using the rule (\ref{3.7}) and does not contribute to 
the Poisson integral. On the contrary the distributional term in the 
interaction part ($\propto m_1^2 m_2$) is well-defined and gives a net 
contribution. Easy terms are obtained from the facts that 
$\Delta (1/r_1^3)=6/r_1^5$ and $\Delta (r_1)=2/r_1$. The difficult point is 
to find the solutions of the two Poisson equations

\begin{mathletters}
\label{6.3}
\begin{eqnarray}
\Delta K_1 &=& 2~\partial^2_{ij}\left({1\over r_2}\right)
\partial^2_{ij}\ln r_1  \ , \label{6.3a} \\
\Delta H_1 &=& 2~\partial^2_{ij}\left({1\over r_1}\right){}_ig_j \ . 
\label{6.3b}
\end{eqnarray}
\end{mathletters}
We use the same notation as in \cite{BDI95} except that we add a subscript 1 
to distinguish a function from its image obtained by the exchange of bodies 1 
and 2. Remarkably, the solutions of (\ref{6.3}) can be written down 
everywhere in space-time under explicit form \cite{Dunpublished,BDI95,N10}:

\begin{mathletters}
\label{6.4}
\begin{eqnarray}
K_1 &=& \left({1\over 2}\Delta-\Delta_1\right)
\left[{\ln r_1\over r_2}\right]\nonumber\\
&+&{1\over 2}\Delta_2\left[{\ln r_{12}\over r_2}\right]
+{r_2\over 2r_{12}^2r_1^2}+{1\over r_{12}^2r_2} \ , \label{6.4a} \\
H_1 &=& {1\over 2}\Delta_1\left[{g\over r_1}+{\ln r_1\over r_{12}}-
\Delta_1\left({r_1+r_{12}\over 2}g \right)\right]\nonumber\\
&+&\partial_i\partial_{2i}\left[{\ln r_{12} \over r_1}+
{\ln r_1\over 2r_{12}}\right]
-{1\over r_1}\partial_{2i}[(\partial_i g)_1]-{r_2\over 2r_1^2r_{12}^2}
 \ . \label{6.4b} 
\end{eqnarray}
\end{mathletters}
They are equivalent to the expressions given by the equations (3.48)-(3.49) 
in \cite{BDI95}. By expanding all derivatives, we come to the completely 
developed forms 

\begin{mathletters}
\label{6.5}
\begin{eqnarray}
K_1 &=& -{1 \over r_2^3} + {1 \over r_2r_{12}^2}-{1 \over r_1^2r_2}+
{r_2 \over 2r_1^2r_{12}^2}+{r_{12}^2 \over 2r_1^2r_2^3}
+{r_1^2 \over 2r_2^3r_{12}^2} \ , \label{6.5a} \\
H_1 &=& -{1 \over 2r_1^3}-{1 \over 4r_{12}^3} - {1 \over 4r_1^2r_{12}}-
{r_2 \over 2 r_1^2r_{12}^2}+{r_2 \over 2r_1^3r_{12}}
+{3r_2^2 \over 4r_1^2r_{12}^3}+{r_2^2 \over 2r_1^3r_{12}^2}-
{r_2^3 \over 2r_1^3r_{12}^3} \label{6.5b}
\end{eqnarray}
\end{mathletters}
($K_2$ and $H_2$ are obtained by exchanging $r_1$ and $r_2$ in the 
right-hand-sides). With the solutions (\ref{6.4})-(\ref{6.5}), we control 
the NC potential at the Newtonian approximation.

Next, we compute the spatial integral of (\ref{6.2}) entering the 0.5PN 
correction in the NC potential. We must evaluate essentially the spatial 
integrals of the two source terms in the right-hand-sides of (\ref{6.3}). We 
proceed like for the integral of $1/r_1r_2$ in (\ref{5.7})-(\ref{5.8}). 
Namely, we integrate over a ball of constant radius ${\cal R}$, and use the 
function $K_1$ to transform the volume integral into a surface integral over 
the sphere $r={\cal R}$: 

\begin{equation}
2\int_{|{\bf x}|\leq {\cal R}} d^3{\bf x}
~\partial^2_{ij}\left({1\over r_2}\right)\partial^2_{ij}\ln r_1 
=\int_{|{\bf x}|\leq {\cal R}} d^3{\bf x}~\Delta K_1
=\int_{r={\cal R}}d\Omega~(r^2\partial_rK_1) \ . \label{6.6}
\end{equation}
From the developed expression of $K_1$ given by (\ref{6.5a}), we get
$K_1=2/rr_{12}^2+O(1/r^2)$, which, when substituted into the surface integral 
in (\ref{6.6}), yields 

\begin{equation}
-{1\over 2\pi}\int_{|{\bf x}|\leq {\cal R}} d^3{\bf x}
~\partial^2_{ij}\left({1\over r_2}\right)\partial^2_{ij}\ln r_1 
={2\over r_{12}^2}+O\left({1\over {\cal R}}\right) \ . \label{6.7}
\end{equation}
As we can see, the integral is finite in the limit ${\cal R}\to \infty$, 
with value

\begin{equation}
-{1\over 2\pi}\int d^3{\bf x}
~\partial^2_{ij}\left({1\over r_2}\right)\partial^2_{ij}\ln r_1 
={2\over r_{12}^2} \ . \label{6.8}
\end{equation}
The same method applied to $H_1$ leads, since $H_1=O(1/r^2)$, to 

\begin{equation}
-{1\over 2\pi}\int d^3{\bf x}~\partial^2_{ij}\left({1\over r_1}\right)
{}_ig_j=0 \ . \label{6.9}
\end{equation}
We must compute now the spatial integral of $1/r_1^5$ [see the first term 
in (\ref{6.2})]. It is clearly infinite because of the divergency at the 
bound ${\bf x}\to {\bf y}_1$. By integrating $1/r^5_1$ from $r_1=\epsilon$ 
up to infinity, we obtain: $\int_{r_1\geq\epsilon} d^3{\bf x}/r_1^5=
2\pi/ \epsilon^2$, which is a pure constant \cite{N11}, cancelled after 
applying the time-derivative in front of the $1/c$-term in (\ref{6.1}). 
The second term in (\ref{6.1}) is therefore

\begin{equation}
{1\over 4\pi c}{d\over dt}\int d^3{\bf x} \left\{ \hat{W}_{ij}^{(\partial V
\partial V)} \partial^2_{ij} V\right\} = -{G^3m_1^2m_2\over 2cr_{12}^3}
(n_{12}v_{12})+O(2)+1\leftrightarrow 2 \ . \label{6.10}
\end{equation}
By summing the various contributions, we find the non-compact potential at 
0.5PN order:

\begin{eqnarray}
\hat{X}^{({\rm NC})} &=& {G^3m_1^3\over 12r_1^3} 
- G^3m_1^2m_2\left\{{1\over 8r_2r_{12}^2}+{1\over 16}K_1+H_1\right\} 
\nonumber\\
&+&{G^3m_1^2m_2\over cr_{12}^2}\left[n_{12}^iv_{12}^j-
{1\over 2}\left(3n_{12}^{ij}-
\delta^{ij}\right)(n_{12}v_{12})\right]\partial^2_{1ij} r_1
\nonumber\\
&-&{G^3m_1^2m_2\over 2cr_{12}^3}(n_{12}v_{12})
+O(2)+1\leftrightarrow 2 \ . \label{6.11}
\end{eqnarray}
Adding the other contributions in ${\hat X}$ we end up with the complete 
expression reported in the Appendix B.

Finally, we give the value of the non-compact potential (\ref{6.11}) at the 
location of body 1. From the Hadamard recipe (\ref{3.6}) we find for the 
functions $K_{1,2}$ and $H_{1,2}$ at point 1:

\begin{mathletters}
\label{6.12}
\begin{eqnarray}
(K_1)_1 &=&{2\over 3r_{12}^3}\ ; \quad (K_2)_1 =0 \ , \label{6.12a}\\
(H_1)_1 &=&{1\over 3r_{12}^3}\ ; \quad (H_2)_1 =-{1\over r_{12}^3} \ , 
\label{6.12b} 
\end{eqnarray}
\end{mathletters}
so that the non-compact potential at point 1 is 

\begin{eqnarray}
({\hat X}^{({\rm NC})})_1 &=& {G^3m_2\over r_{12}^3}\left[-{1\over 2}m_1^2+
m_1m_2+{1\over 12}m_2^2\right]\nonumber\\
&+&{G^3m_1m_2 \over 2cr_{12}^3} (-m_1+m_2)(n_{12}v_{12})+O(2) \ . \label{6.13}
\end{eqnarray}

\section{The 2.5PN metric of binary systems}

From the results of the previous sections, we are in the position to write 
down the 2.5PN harmonic-coordinate metric generated by two point-like 
particles as a function of the coordinate position ${\bf x}$ and a functional 
of the coordinate positions and velocities of the particles 
${\bf y}_{1,2}(t),{\bf v}_{1,2}(t)$ (where $t=$const is the 
harmonic-coordinate slicing):

\begin{equation}
g_{\mu\nu}({\bf x},t) = g_{\mu\nu}[{\bf x}; {\bf y}_1(t),{\bf y}_2(t); 
{\bf v}_1(t),{\bf v}_2(t)] \ . \label{7.1}
\end{equation}
The metric is given by (\ref{2.5}) in which we insert the expressions of the 
potentials as listed in Appendix B. After combining together identical terms 
we obtain \cite{N12}:

\begin{mathletters}
\label{7.2}
\begin{eqnarray}
g_{00}+1 & = &
\frac{2 G m_1}{c^2 r_1}+
\frac{1}{c^4} 
	\left[\frac{G m_1}{r_1} \left(-(n_1v_1)^2+4 v_1^2 \right)
       -2 \frac{G^2 m_1^2}{r_1^2}\right. \nonumber \\ & & 
	\qquad \qquad \quad+
	\left. G^2 m_1 m_2 \left(-\frac{2}{r_1 r_2}-
				\frac{r_1}{2 r_{12}^3}+
				\frac{r_1^2}{2 r_2 r_{12}^3}-
				\frac{5}{2 r_2 r_{12}}\right)\right] 
+\frac{4 G^2 m_1 m_2}{3 c^5 r_{12}^2} (n_{12}v_{12})
\nonumber \\&+&
\frac{1}{c^6} \left[
\frac{G m_1}{r_1} \left(\frac{3}{4} (n_1v_1)^4-3 (n_1v_1)^2 v_1^2+	
							4 v_1^4\right)+
\frac{G^2 m_1^2}{r_1^2} \left(3 (n_1v_1)^2-v_1^2\right) 
				+2 \frac{G^3 m_1^3}{r_1^3}\right. 
				\nonumber \\ &+&
G^2 m_1 m_2 \left(v_1^2 \left(\frac{3 r_1^3}{8 r_{12}^5}-  
	         \frac{3 r_1^2 r_2}{8 r_{12}^5}-
		 \frac{3 r_1 r_2^2}{8 r_{12}^5}+ 
   		 \frac{3 r_2^3}{8 r_{12}^5}-\frac{37 r_1}{8 r_{12}^3}+
		 \frac{r_1^2}{r_2 r_{12}^3}+\frac{3 r_2}{8 r_{12}^3} 
\right. \right. \nonumber
		 \\ & & \left.
		 \qquad \qquad \quad +
		 \frac{2 r_2^2}{r_1 r_{12}^3}+\frac{6}{r_1 r_{12}}- 
   		 \frac{5}{r_2 r_{12}}-\frac{8r_{12}}{r_1r_2S}
		+\frac{16}{r_{12} S}\right)\nonumber \\  & &
		 		\qquad \quad+
	         (v_1v_2) \left(\frac{8}{r_1 r_2}-\frac{3 r_1^3}{4 r_{12}^5}+
		 \frac{3 r_1^2 r_2}{4 r_{12}^5}+ 
  		 \frac{13 r_1}{4 r_{12}^3}-\frac{2 r_1^2}{r_2 r_{12}^3}-
		 \frac{6}{r_1 r_{12}}-
		 \frac{16}{r_1 S}- \frac{12}{r_{12} S}\right) \nonumber \\ & &
		    		\qquad \quad+
	   (n_{12}v_1)^2 \left(-\frac{15 r_1^3}{8 r_{12}^5}+
			       \frac{15 r_1^2 r_2}{8 r_{12}^5}+
  		    	       \frac{15 r_1 r_2^2}{8 r_{12}^5}-
			       \frac{15 r_2^3}{8 r_{12}^5}+
  		    	       \frac{57 r_1}{8 r_{12}^3}-
			       \frac{3 r_1^2}{4 r_2 r_{12}^3}-
  		    	       \frac{33 r_2}{8 r_{12}^3}
			       \right. \nonumber \\ & & \left. 
			       \qquad \qquad \qquad \quad+
			       \frac{7}{4 r_2 r_{12}} -
			       \frac{16}{S^2}-\frac{16}{r_{12} S}\right)
			\nonumber \\ & & \quad \qquad +
	   (n_{12}v_1) (n_{12}v_2) \left(\frac{15 r_1^3}{4 r_{12}^5}-
			\frac{15 r_1^2 r_2}{4 r_{12}^5}-
  			\frac{9 r_1}{4 r_{12}^3}+\frac{12}{S^2}+
			\frac{12}{r_{12} S}\right) \nonumber
			\\ & & \quad \qquad+
	   (n_1v_1)^2 \left(\frac{2}{r_1 r_2}-\frac{r_1}{4 r_{12}^3}-
			    \frac{3 r_2^2}{4 r_1 r_{12}^3}+
  		   	    \frac{7}{4 r_1 r_{12}}-\frac{8}{S^2}-
			    \frac{8}{r_1 S}\right) \nonumber
			\\ & & \quad \qquad+
	   (n_1v_1) (n_1v_2) \left(\frac{r_1}{r_{12}^3}+\frac{16}{S^2}+
				   \frac{16}{r_1 S}\right)-
	   (n_1v_2)^2 \left(\frac{8}{S^2}+\frac{8}{r_1 S}\right) \nonumber
	   \\ & & \qquad \quad+
	   (n_{12}v_1) (n_1v_1) \left(-\frac{3 r_1^2}{r_{12}^4}+
				\frac{3 r_2^2}{2 r_{12}^4}+
				\frac{3}{2 r_{12}^2}+\frac{16}{S^2}\right)+
	   \frac{16 (n_1v_2) (n_2v_1)}{S^2}
			\nonumber \\ & & \qquad \quad +
	   (n_{12}v_2) (n_1v_1) \left(\frac{3 r_1^2}{r_{12}^4}-
				      \frac{3 r_2^2}{2 r_{12}^4}+
				      \frac{13}{2 r_{12}^2}-
				      \frac{40}{S^2}\right)-
		\frac{12(n_1v_1) (n_2v_2)}{S^2} \nonumber
			\\ & & \left. \qquad \quad +
	   (n_{12}v_1) (n_1v_2) \left(\frac{3 r_1^2}{2 r_{12}^4}+
				      \frac{4}{r_{12}^2}+
				      \frac{16}{S^2} \right)+
	   (n_{12}v_2) (n_1v_2) \left(\frac{-3 r_1^2}{2 r_{12}^4}-
				      \frac{3}{r_{12}^2}+\frac{16}{S^2}\right)
					\right)
			\nonumber \\ &+&
G^3 m_1^2 m_2 \left(\frac{4}{r_1^3}+\frac{1}{2 r_2^3}+\frac{9}{2 r_1^2 r_2}-
	       \frac{r_1^3}{4 r_{12}^6}+\frac{3 r_1^4}{16 r_2 r_{12}^6}-
	       \frac{r_1^2 r_2}{8 r_{12}^6}+
  	       \frac{r_1 r_2^2}{4 r_{12}^6}-
	       \frac{r_2^3}{16 r_{12}^6}+
	       \frac{5 r_1}{4 r_{12}^4}
			\right. \nonumber \\ & & \qquad \ 
		-\frac{23 r_1^2}{8 r_2 r_{12}^4}+
	       \frac{43 r_2}{8 r_{12}^4}-
  	       \frac{5 r_2^2}{2 r_1 r_{12}^4}-\frac{3}{r_{12}^3}+
	       \frac{3 r_1}{r_2 r_{12}^3}+\frac{r_2}{r_1 r_{12}^3}-
	       \frac{5 r_2^2}{r_1^2 r_{12}^3}+
  	       \frac{4 r_2^3}{r_1^3 r_{12}^3}+\frac{3}{2 r_1 r_{12}^2}
	\nonumber \\ & & \left. \left. \qquad \ -
  	       \frac{r_1^2}{4 r_2^3 r_{12}^2} +
	       \frac{3}{16 r_2 r_{12}^2}+
  	       \frac{15 r_2}{4 r_1^2 r_{12}^2}-\frac{4 r_2^2}{r_1^3 r_{12}^2}+
  	       \frac{5}{r_1^2 r_{12}}+\frac{5}{r_1 r_2 r_{12}}-
	       \frac{4 r_2}{r_1^3 r_{12}}-
	       \frac{r_{12}^2}{4 r_1^2 r_2^3}\right)\right] \nonumber \\ &+&
\frac{1}{c^7} \left[
G^2 m_1 m_2
   \left((n_{12}v_{12})^2 (n_1v_1) \left(-\frac{8 r_1^3}{r_{12}^5}-
	 \frac{16 r_1}{r_{12}^3}\right)+
     	(n_{12}v_{12})^2 (n_1v_2) \left(\frac{8 r_1^3}{r_{12}^5}+
	\frac{5 r_1}{r_{12}^3}\right) \right. \right.
\nonumber \\ & & \qquad \qquad \quad +
     	(n_{12}v_{12})^3 \left(-\frac{7 r_1^4}{2 r_{12}^6}+
	\frac{7 r_1^2 r_2^2}{2 r_{12}^6}-
        \frac{11 r_1^2}{r_{12}^4}-\frac{37}{4 r_{12}^2}\right)
\nonumber \\ & & \qquad \qquad \quad +
     	(n_{12}v_1) (n_{12}v_{12})^2 \left(\frac{20 r_1^2}{r_{12}^4}-
	\frac{11}{2 r_{12}^2}\right)-
     	4 (n_{12}v_{12}) (n_1v_1)^2 \frac{r_1^2}{r_{12}^4}
\nonumber \\ & & \qquad \qquad \quad
     	+4 (n_{12}v_{12}) (n_1v_1) (n_1v_2)\frac{r_1^2}{r_{12}^4}+
     	(n_{12}v_1)^2 (n_1v_1)\frac{r_1}{r_{12}^3}
\nonumber \\ & & \qquad \qquad \quad+
     	22 (n_{12}v_1) (n_{12}v_{12}) (n_1v_1)\frac{r_1}{r_{12}^3}-
     	(n_{12}v_1)^2 (n_1v_2)\frac{r_1}{r_{12}^3}
\nonumber \\ & & \qquad \qquad \quad+
	4 (n_{12}v_1) (n_{12}v_{12}) (n_1v_2)\frac{r_1}{r_{12}^3}+
     	11 (n_{12}v_1)^2 (n_{12}v_{12})\frac{1}{2 r_{12}^2}
\nonumber \\ & & \qquad \qquad \quad+
     	(n_1v_2) v_{12}^2 \left(-\frac{8 r_1^3}{5 r_{12}^5}-
		\frac{2 r_1}{3 r_{12}^3}\right)+
     	(n_1v_1) v_{12}^2 \left(\frac{8 r_1^3}{5 r_{12}^5}+
		\frac{11 r_1}{3 r_{12}^3}\right)
\nonumber \\ & & \qquad \qquad \quad -
     	(n_{12}v_1) v_{12}^2 \left(\frac{4 r_1^2}{r_{12}^4}+
	\frac{5}{2 r_{12}^2}\right)-
     	(n_{12}v_{12}) v_1^2 \left(\frac{12 r_1^2}{r_{12}^4}+
		\frac{5}{2 r_{12}^2}\right)
\nonumber \\ & & \qquad \qquad \quad+
     	(n_{12}v_{12}) v_{12}^2 \left(\frac{3 r_1^4}{2 r_{12}^6}-
	\frac{3 r_1^2 r_2^2}{2 r_{12}^6}+
       \frac{7 r_1^2}{r_{12}^4}+\frac{27}{4 r_{12}^2}\right)
\nonumber \\ & & \qquad \qquad \quad-
     	29 (n_1v_1) v_1^2\frac{r_1}{3 r_{12}^3}+
	(n_1v_2) v_1^2\frac{r_1}{r_{12}^3}+
     	5 (n_{12}v_1) v_1^2\frac{1}{r_{12}^2}
\nonumber \\ & & \qquad \qquad \quad+
     	(n_{12}v_{12}) (v_1v_2) \left(\frac{12 r_1^2}{r_{12}^4}+
		\frac{3}{r_{12}^2}\right)+
     	8 (n_1v_1) (v_1v_2)\frac{r_1}{r_{12}^3}
\nonumber \\ & & \left. \qquad \quad \qquad+
	2 (n_1v_2) (v_1v_2)\frac{r_1}{3 r_{12}^3}-
     	5 (n_{12}v_1) (v_1v_2)\frac{1}{r_{12}^2}\right)
\nonumber \\ &+ &
G^3 m_1^2 m_2 \left((n_1v_{12}) \left(-\frac{8 r_1^3}{15 r_{12}^6}+
		\frac{8 r_1 r_2^2}{15 r_{12}^6}-
        \frac{16 r_1}{3 r_{12}^4}+\frac{8}{r_{12}^3}-
	\frac{8 r_2^2}{r_1^2 r_{12}^3}+
  	\frac{8}{r_1^2 r_{12}}\right) \right.
\nonumber \\ & & \qquad \quad +(n_{12}v_1) 
      	\left(-\frac{4 r_1^2}{3 r_{12}^5}+
		\frac{4 r_2^2}{3 r_{12}^5}+\frac{20}{3 r_{12}^3}\right)+
     	8 (n_1v_1)\frac{r_1}{3 r_{12}^4}
\nonumber \\ & & \qquad \quad +
     (n_{12}v_{12}) \left(\frac{3}{r_1^3}-\frac{4 r_1^2}{3 r_{12}^5}+
		\frac{68 r_2^2}{15 r_{12}^5}+
       \frac{3 r_1}{r_{12}^4}-\frac{6 r_2^2}{r_1 r_{12}^4}+
	\frac{3 r_2^4}{r_1^3 r_{12}^4}-
       \frac{76}{3 r_{12}^3} \right.
\nonumber \\ & &  \left. \left. \left. \quad \qquad 
\qquad \qquad +\frac{2}{r_1 r_{12}^2}-
	\frac{6 r_2^2}{r_1^3 r_{12}^2}\right)\right)
\right]+O(8)+1 \leftrightarrow 2     \ , \label{7.2a}
\\ \nonumber\\ 
g_{0i} & = &
-4 \frac{G m_1}{c^3 r_1} v_1^i+
\frac{1}{c^5} \left[
n_1^i \left(-\frac{G^2 m_1^2}{r_1^2} (n_1v_1)+
		\frac{G^2 m_1 m_2}{S^2} \left(-16 (n_{12}v_1)+
	12 (n_{12}v_2)
			\right. \right. \right. \nonumber \\ & & \left. 
			\frac{}{}
			\left. \qquad \qquad \qquad 
	-16 (n_2v_1) +12 (n_2v_2) \right) \right) \nonumber \\
			& & \qquad \qquad +
n_{12}^i G^2 m_1 m_2 \left(-6 (n_{12}v_{12}) \frac{r_1}{r_{12}^3}-
	4 (n_1v_1)\frac{1}{r_{12}^2}+12 (n_1v_1)\frac{1}{S^2} \right.
			\nonumber \\ & & \left. \qquad \qquad \qquad \qquad
			\qquad \qquad \qquad-
	16 (n_1v_2)\frac{1}{S^2}+
	4 (n_{12}v_1)\frac{1}{S} \left(\frac{1}{S}+
		\frac{1}{r_{12}}\right) \right) \nonumber \\ & & 
			\qquad \qquad +
v_1^i \left(\frac{G m_1}{r_1} \left(2 (n_1v_1)^2-4 v_1^2\right)+
		\frac{G^2 m_1^2}{r_1^2}+
	G^2 m_1 m_2 \left(\frac{3 r_1}{r_{12}^3}-\frac{2 r_2}{r_{12}^3}\right)
			\right. \nonumber \\ & & \left. \left. 
			\qquad \qquad \qquad \qquad  +
	G^2 m_1 m_2  \left(-\frac{r_2^2}{r_1 r_{12}^3}  -\frac{3}{r_1 r_{12}}+
  		\frac{8}{r_2 r_{12}}-\frac{4}{r_{12} S}\right)\right)\right]
			\nonumber \\ &+&
\frac{1}{c^6} \left[
n_{12}^i \left(G^2 m_1 m_2 \left(-10 (n_{12}v_{12})^2 \frac{r_1^2}{r_{12}^4}-
	12 (n_{12}v_{12}) (n_1v_1) \frac{r_1}{r_{12}^3}+
	2 v_{12}^2 \frac{r_1^2}{r_{12}^4}-4 \frac{v_1^2}{r_{12}^2}\right)
			\right. \right. \nonumber \\ & & \left. \qquad \quad+
        G^3 m_1^2 m_2 \left(\frac{2 r_1^2}{3 r_{12}^5}-
		\frac{2 r_2^2}{3 r_{12}^5}-
		\frac{2}{r_{12}^3}\right)\right)
			\nonumber \\ 
			& & \qquad +
v_1^i \frac{G^2 m_1 m_2}{r_{12}^3} \left(\frac{16 (n_1v_{12}) r_1}{3}-
4 (n_{12}v_2) r_{12} \right)
			\nonumber \\ & & \qquad \left. +
v_{12}^i \frac{G^2 m_1 m_2}{r_{12}^2} \left(-2 (n_{12}v_1)+
	6 (n_{12}v_{12}) \frac{r_1^2}{r_{12}^2}\right) \right]+O(7)+
1 \leftrightarrow 2             \ , \label{7.2b}
\\ \nonumber\\
g_{ij} -\delta_{ij} & = &
2 \frac{G m_1}{c^2 r_1} \delta^{ij}+
\frac{1}{c^4} \left[
\delta^{ij} \left(
		-\frac{G m_1}{r_1} (n_1v_1)^2+\frac{G^2 m_1^2}{r_1^2}
			\right. \right.
			\nonumber \\ & & \left. \qquad \qquad 
			\qquad \qquad   +
		G^2 m_1 m_2 \left(\frac{2}{r_1 r_2}-\frac{r_1}{2 r_{12}^3}+
		\frac{r_1^2}{2 r_2 r_{12}^3}-\frac{5}{2 r_1 r_{12}}+  
  		\frac{4}{r_{12} S}\right)\right) 
			\nonumber \\ & & \qquad \qquad \qquad +
4 \frac{G m_1}{r_1} v_1^i v_1^j+
\frac{G^2 m_1^2}{r_1^2} n_1^i n_1^j -
4 G^2 m_1 m_2 n_{12}^i n_{12}^j \left(\frac{1}{S^2}+
		\frac{1}{r_{12} S}\right)
			\nonumber \\ & & \left.
			\qquad \qquad \qquad+
\frac{4 G^2 m_1 m_2}{S^2} \left(n_1^{(i} n_2^{j)}+
	2n_1^{(i} n_{12}^{j)} \right)\right]
			\nonumber \\ &+&
\frac{G^2 m_1 m_2}{c^5r_{12}^2} \left(-\frac{2}{3}(n_{12}v_{12})\delta^{ij}-
6 (n_{12}v_{12}) n_{12}^i n_{12}^j+ 8 n_{12}^{(i} v_{12}^{j)}\right)+O(6)+
1 \leftrightarrow 2    \ , \label{7.2c}
\end{eqnarray}
\end{mathletters}
(where we recall $S=r_1+r_2+r_{12}$). The harmonic-coordinate conditions 
satisfied by (\ref{7.2}) read, with this order of approximation: 
$\partial_\nu [(-g)^{1/2} g^{0\nu}]=O(7)$ and 
$\partial_\nu [(-g)^{1/2} g^{i\nu}]=O(6)$.

When applied to post-Newtonian initial conditions for the numerical evolution 
of two compact objects \cite{SSSTWW97,BCTh98} (provided that the initial 
spatial numerical grid does not extend outside the binary's near zone 
\cite{N2}), the lengthy expressions (\ref{7.2}) become a little simpler as 
the orbit can be considered as circular with a good approximation. In this 
case we have $(n_{12}v_1)=O(5)=(n_{12}v_2)$, the remainder $O(5)$ 
corresponding to radiation-reaction effects. Obviously, all the resulting 
$O(5)$'s can be neglected since they yield terms falling into the 
uncontrolled remainders of (\ref{7.2}). If in addition we are working in a 
mass centered frame, then ${\bf y}_1=X_2{\bf y}_{12}+O(4)$ and 
${\bf y}_2=-X_1{\bf y}_{12}+O(4)$, where $X_1=m_1/m$, $X_2=m_2/m$, 
$m=m_1+m_2$. All the remainders $O(4)$ become negligible after insertion 
in (\ref{7.2}). Thus, for instance: $v_1^2=X_2^2v_{12}^2+O(4)$, and 

\begin{mathletters}
\label{7.3}
\begin{eqnarray}
(n_1v_1)&=&X_2{r\over r_1}(nv_{12})+O(4) \ , \label{7.3a}\\
(n_1v_2)&=&-X_1{r\over r_1}(nv_{12})+O(4)  \label{7.3b}
\end{eqnarray}
\end{mathletters}
(plus the same formulas with 1 and 2 exchanged). We denote by 
${\bf n}= {\bf x}/r$ and $r=|{\bf x}|$ the direction and the distance from 
the center of mass; $r$ depends on the two individual distances $r_{1,2}$ 
through the relation

\begin{equation} 
r^2=X_1r_1^2+X_2r_2^2-X_1X_2r_{12}^2+O(4)   \ . \label{7.4}
\end{equation}
The magnitude of the relative velocity is 
$v_{12}=r_{12}~\omega_{\rm 2PN}+O(6)$, where $\omega_{\rm 2PN}$ denotes the 
orbital frequency of the circular motion at 2PN order, and is given by (8.6) 
below \cite{N13}.
 
We provide also the values of the metric coefficients (\ref{7.2}) computed 
at body 1 (since these might also be needed in the problem of binary 
coalescence), i.e.

\begin{equation}
(g_{\mu\nu})_1(t) = g_{\mu\nu}[{\bf y}_1(t); {\bf y}_1(t),{\bf y}_2(t); 
{\bf v}_1(t),{\bf v}_2(t)] \ , \label{7.5}
\end{equation}
where the limit ${\bf x}\to {\bf y}_1$ is understood in the sense of  
(\ref{3.6}). Directly from (\ref{7.2}), or using the expressions of the 
potentials at 1 as given in Appendix B, we get:

\begin{mathletters}
\label{7.6}
\begin{eqnarray}
(g_{00})_1 &=&-1+2 \frac{G m_2}{c^2r_{12}} 
+\frac{G m_2}{c^4 r_{12}} \left(4 v_2^2-(n_{12}v_2)^2-3\frac{G m_1}{r_{12}}
- 2 \frac{G m_2}{r_{12}} \right) \nonumber \\ 
&+&\frac{8G^2 m_1 m_2}{3c^5 r_{12}^2}(n_{12}v_{12})
+ \frac{G m_2}{c^6 r_{12}} \left( \frac{3}{4} (n_{12}v_2)^4- 
3 (n_{12}v_2)^2 v_2^2 + 4 v_2^4 \right) \nonumber \\ &+&
\frac{G^2 m_1 m_2}{c^6 r_{12}^2} \left(-\frac{87}{4} (n_{12}v_1)^2 + 
\frac{47}{2}(n_{12}v_1) (n_{12}v_2)- 
  \frac{55}{4} (n_{12}v_2)^2 + \frac{23}{4} v_1^2 - 
  \frac{39}{2} (v_1v_2) \right) \nonumber \\
&+& \frac{47}{4} \frac{G^2 m_1 m_2}{c^6 r_{12}^2} v_2^2+
\frac{G m_2}{c^6 r_{12}} \left\{ \frac{G m_2}{r_{12}} 
\left[3 (n_{12}v_2)^2 - v_2^2 \right]
-\frac{G^2 m_1^2}{r_{12}^2} + \frac{17}{2} \frac{G^2 m_1 m_2}{r_{12}^2} + 
  2 \frac{G^2 m_2^2}{r_{12}^2}  \right\} \nonumber \\ 
&+& \frac{G^2 m_1 m_2}{c^7 r_{12}^2} \left\{ -20 (n_{12}v_1)^3 + 
40 (n_{12}v_1)^2 (n_{12}v_2) - 36 (n_{12}v_1) (n_{12}v_2)^2 + 
  16 (n_{12}v_2)^3 \right. \nonumber 
\\ & & \qquad \qquad+\frac{296}{15} (n_{12}v_1) v_1^2  - 
\frac{116}{15} (n_{12}v_2) v_1^2 - 
  \frac{104}{5} (n_{12}v_1) (v_1v_2) + \frac{232}{15} (n_{12}v_2) (v_1v_2) 
\nonumber \\ & & \qquad \qquad+ 
 \frac{56}{15} (n_{12}v_1) v_2^2 - \frac{52}{5} (n_{12}v_2) v_2^2
+\frac{G m_1}{r_{12}} \left(-\frac{64}{5} (n_{12}v_1)+
\frac{104}{5} (n_{12}v_2) \right)
\nonumber \\ & & \qquad \qquad \left.
+\frac{G m_2}{r_{12}} \left(-\frac{144}{5}(n_{12}v_1)+
\frac{392}{15}(n_{12}v_2)\right)
\right\}+O(8)   \ , \label{7.6a}\\
(g_{0i})_1 &=&-4 \frac{G m_2}{c^3 r_{12}} v_2^i
+\frac{G m_2}{c^5 r_{12}}\left\{n_{12}^i 
\left[ \frac{G m_1}{r_{12}} \left(10 (n_{12}v_1) + 2 (n_{12}v_2)\right)
-\frac{G m_2}{r_{12}} (n_{12}v_2)\right]
\right\} \nonumber \\ 
&+& \frac{G m_2}{c^5 r_{12}} \left\{ 4 \frac{G m_1}{r_{12}} v_1^i 
+v_2^i \left(2 (n_{12}v_2)^2 - 
4 v_2^2- 2 \frac{G m_1}{r_{12}} + 
\frac{G m_2}{r_{12}} \right) \right\}  \nonumber \\ 
&+&\frac{G^2 m_1 m_2}{c^6 r_{12}^2} \left\{n_{12}^i \left(10 (n_{12}v_1)^2 - 
8 (n_{12}v_1) (n_{12}v_2) 
- 2 (n_{12}v_2)^2 - 6 v_1^2  \right. \right. 
\nonumber \\ & & \left. \qquad \qquad \qquad + 4 (v_1v_2) + 2 v_2^2- 
\frac{8}{3} \frac{G m_1}{r_{12}} + \frac{4}{3} \frac{G m_2}{r_{12}} \right)
\nonumber \\ & &   \left. \qquad \qquad -8 (n_{12}v_1) v_1^i
+v_2^i \left(\frac{20}{3} (n_{12}v_1) + 
\frac{4}{3} (n_{12}v_2)\right) \right\} +O(7) \ , \label{7.6b} \\
(g_{ij})_1 &=& \delta^{ij}
+2 \frac{G m_2}{c^2 r_{12}} \delta^{ij}
+\frac{G m_2}{c^4 r_{12}} \delta^{ij} \left(-(n_{12}v_2)^2 +
 \frac{G m_1}{r_{12}} + \frac{G m_2}{r_{12}} \right) \nonumber \\ 
&+&\frac{G m_2}{c^4 r_{12}} \left\{n_{12}^{ij} 
\left(-8 \frac{G m_1}{r_{12}} + 
\frac{G m_2}{r_{12}}\right) + 4 v_2^{ij}\right\} 
-\frac{4 G^2 m_1 m_2}{3c^5 r_{12}^2}
\delta^{ij} (n_{12}v_{12}) \nonumber \\ 
&+&\frac{G^2 m_1 m_2}{c^5 r_{12}^2} \left\{ 
-12 n_{12}^{ij}(n_{12}v_{12})+
16 n_{12}^{(i} v_{12}^{j)}\right\} +O(6) \ . \label{7.6c}
\end{eqnarray}
\end{mathletters}
Drastic simplifications occur in the case where the orbit is circular. 

\section{The 2.5PN equations of motion of binary systems}

The motion of body 1 under the gravitational influence of body 2 is simply 
the geodesic motion taking place in the post-Newtonian space-time 
(\ref{7.2}). Now we write the geodesic equation of body 1 in the 
Newtonian-like form

\begin{equation}
{d{\cal P}_1^i\over dt} = {\cal F}_1^i   \ . \label{8.1}
\end{equation}
The ``linear momentum'' vector and ``gravitational force'' (per unit of mass) 
are defined by

\begin{equation}
{\cal P}_1^i \equiv \left({v_1^\mu g_{i\mu}\over 
\sqrt{-g_{\rho\sigma} {\displaystyle \frac{v_1^\rho v_1^\sigma}{c^2}}}}
\right)_1 
\qquad\quad ; \qquad\quad
{\cal F}_1^i \equiv {1\over 2} 
\left( {v_1^\mu v_1^\nu \partial_i g_{\mu\nu} \over 
\sqrt{-g_{\rho\sigma} {\displaystyle \frac{v_1^\rho v_1^\sigma}{c^2}}}}  
\right)_1 \ . \label{8.2}
\end{equation}
The above quantities are computed using the 
regularization (\ref{3.6}) which is a crucial ingredient of our point-mass 
model.

Inserting the 2.5PN metric (\ref{2.5}) into (\ref{8.2}), we obtain 

\begin{mathletters}
\label{8.3}
\begin{eqnarray}
{\cal P}_1^i = v_1^i&+&{1\over c^2}\left[-4(V_i)_1+3(V)_1v_1^i+
{1\over 2}v_1^2v_1^i\right] \nonumber\\
&+&{1\over c^4}\left[-8({\hat R}_i)_1+{9\over 2}(V^2)_1v_1^i+
4({\hat W}_{ij})_1v_1^j-4(VV_i)_1 
\right.\nonumber\\
&+&\left.{7\over 2}(V)_1v_1^2v_1^i-2v_1^2(V_i)_1-4v_1^iv_1^j(V_j)_1+
{3\over 8}v_1^iv_1^4\right]+O(6) \ , \label{8.3a} \\
\nonumber \\
{\cal F}_1^i = (\partial_iV)_1&+&{1\over c^2}\left[-(V\partial_iV)_1+
{3\over 2}v_1^2(\partial_iV)_1- 4v_1^j(\partial_iV_j)_1\right] \nonumber\\
&+&{1\over c^4}\left[4(\partial_i{\hat X})_1+8(V_j\partial_iV_j)_1-
8v_1^j(\partial_i{\hat R}_j)_1+{9\over 
2}v_1^2(V\partial_iV)_1 \right.\nonumber\\
&+& \left.2v_1^jv_1^k(\partial_i{\hat W}_{jk})_1-2v_1^2v_1^j(\partial_iV_j)_1+
{7\over 8}v_1^4(\partial_iV)_1 +{1\over 2}(V^2 \partial_iV)_1\right.\nonumber\\
&-&\left. 4v_1^j(V_j\partial_iV)_1-4v_1^j(V\partial_iV_j)_1\right]+O(6)
\ . \label{8.3b}
\end{eqnarray}
\end{mathletters}
Next we replace into these expressions all the potentials and their gradients 
computed at point 1 as given in Appendix B (to this order the Hadamard partie 
finie is ``distributive'' \cite{N6}), and get both ${\cal P}_1^i$ and 
${\cal F}_1^i$ in terms of the relative separation $y_{12}^i=r_{12}n_{12}^i$ 
and individual velocities $v_{1,2}^i$ [alternatively we can obtain 
${\cal P}^i_1$ and ${\cal F}^i_1$ directly from (\ref{7.2})]. Then, we 
compute the time-derivative of ${\cal P}_1^i$, and  order-reduce all the 
resulting accelerations (which appear at orders 1PN or 2PN) by means of 
the 1.5PN equations of motion given by (\ref{3.11}). After insertion in 
(\ref{8.1}) and simplification, we end with the 2.5PN acceleration of body 1:

\begin{eqnarray}
{dv_1^i\over dt} = &-& {Gm_2\over r_{12}^2} n_{12}^i + 
{Gm_2\over r_{12}^2c^2}\biggl\{v_{12}^i \left[4(n_{12}v_1) -
3(n_{12}v_2)\right]  \nonumber\\
 &&\quad  +n_{12}^i \left[ -v^2_1 - 2v^2_2 + 4(v_1v_2)
  + {3\over 2} (n_{12}v_2)^2 + 5{Gm_1\over r_{12}} + 4{Gm_2\over r_{12}} 
\right]\biggr\}\nonumber\\
  &+&{Gm_2\over r_{12}^2c^4} n_{12}^i \biggl\{\left[ -2v^4_2 + 
4v^2_2 (v_1v_2) - 2(v_1v_2)^2  + {3\over 2} v^2_1 (n_{12}v_2)^2 
 +{9\over 2} v^2_2 (n_{12}v_2)^2 \right.\nonumber\\
&&\quad \left. -6(v_1v_2) (n_{12}v_2)^2
  - {15\over 8} (n_{12}v_2)^4 \right]    \nonumber\\
&&\quad +{Gm_1\over r_{12}}\left[ -{15\over 4} v^2_1 +{5\over 4} v^2_2
    -{5\over 2} (v_1v_2) \right.\nonumber\\
&&\quad \left.+ {39\over 2} (n_{12}v_1)^2
    -39(n_{12}v_1)(n_{12}v_2)+{17\over 2}(n_{12}v_2)^2 \right]\nonumber\\
&&\quad +{Gm_2\over r_{12}}\left[ 4 v^2_2 - 8(v_1v_2)+ 2(n_{12}v_1)^2
  - 4(n_{12}v_1)(n_{12}v_2) - 6(n_{12}v_2)^2\right]\nonumber \\
&&\quad +{G^2\over r_{12}^2}\left[ -{57\over 4}m^2_1 - 9m^2_2
   - {69\over 2} m_1m_2 \right]\biggr\}\nonumber\\
&+& {Gm_2\over r_{12}^2c^4}v_{12}^i \biggl\{ v^2_1(n_{12}v_2)+
4v^2_2(n_{12}v_1) -5v^2_2(n_{12}v_2)
    -4(v_1v_2)(n_{12}v_1)\nonumber\\
&&\quad + 4(v_1v_2)(n_{12}v_2) -6(n_{12}v_1)(n_{12}v_2)^2
    + {9\over 2} (n_{12}v_2)^3 \nonumber\\
&&\left. \quad +{Gm_1\over r_{12}} \left[ -{63\over 4}(n_{12}v_1) + {55\over 4}
  (n_{12}v_2)\right] 
+{Gm_2\over r_{12}}\left[-2(n_{12}v_1)-2(n_{12}v_2)\right] \right\}
    \nonumber\\
&+&{4G^2m_1m_2\over 5c^5r_{12}^3}\biggr\{ n_{12}^i (n_{12}v_{12}) 
\left[-6{Gm_1\over r_{12}}+{52\over 3}{Gm_2\over 
r_{12}}+3v_{12}^2\right]\nonumber\\
&&\quad +v_{12}^i \left[2{Gm_1\over r_{12}}-8{Gm_2\over r_{12}}-
v_{12}^2\right]\biggl\}+O(6) \ . \label{8.4}
\end{eqnarray}
We find perfect agreement with the Damour-Deruelle 
\cite{DD81a,Dthese,D82,D83a} equations of motion. To emphasize the strength 
of this agreement we recall that the method employed in the present paper 
differs in many respects from the one originally used in 
\cite{DD81a,Dthese,D82,D83a} (see the discussion in the introduction). In the 
case of circular orbits, the equations reduce to

\begin{equation}
 {dv_{12}^i\over dt} = -\omega^2_{\rm 2PN} y_{12}^i-
{32G^3m^3\nu\over 5c^5r_{12}^4}v_{12}^i + O(6) \ . \label{8.5}
\end{equation}
The second term represents the standard damping force, while the orbital 
frequency $\omega_{\rm 2PN}$ is the frequency of the exact circular motion 
at 2PN order, related to the harmonic-coordinate separation $r_{12}$ by

\begin{equation}
 \omega^2_{\rm 2PN} \equiv {Gm\over r_{12}^3} \left[ 1+(-3+\nu) \gamma +
 \left( 6 +{41\over 4} \nu +\nu^2 \right) \gamma^2 \right] \ . \label{8.6}
\end{equation}
Our notation is: $m=m_1+m_2$, $\nu=m_1m_2/m^2$ ($=X_1X_2$ in the notation of 
the previous section) and $\gamma=Gm/r_{12}c^2$.
 
\appendix
\section{Derivation of the 2.5PN fluid metric}

The derivation follows almost immediately from the results established in the 
section II.A of \cite{B95}. The Einstein field equations in harmonic 
coordinates are written as

\begin{mathletters}
\label{A.1}
\begin{eqnarray}
 \partial_\nu h^{\mu\nu} &=& 0 \ , \label{A.1a}\\
\Box h^{\mu\nu} &=& {16\pi G\over c^4} |g| T^{\mu\nu} +
\Lambda^{\mu\nu} (h) \ , \label{A.1b}
\end{eqnarray}
\end{mathletters}
where $\Box =\eta^{\mu\nu} \partial_\mu\partial_\nu$ is the flat
d'Alembertian operator [$\eta^{\mu\nu} = {\rm diag} (-1,1,1,1)$],  
$h^{\mu\nu}\equiv\sqrt{-g} g^{\mu\nu}- \eta^{\mu\nu}$, and  
$\Lambda^{\mu\nu}$ denotes the gravitational source term which is at least 
quadratic in $h$ and its space-time derivatives (see \cite{B95} for the 
expressions of the quadratic and cubic parts of $\Lambda^{\mu\nu}$).
From \cite{B95,N4}, we have, to order 1PN,

\begin{mathletters}
\label{A.2}
\begin{eqnarray}
h^{00} &=& -{4\over c^2} V -{2\over c^4} \left({\hat W}_{kk}+4V^2\right) +
O(6) \ , \label{A.2a} \\
h^{0i} &=& -{4\over c^3} V_i +O(5) \ , \label{A.2b} \\
h^{ij} &=& -{4\over c^4} \left[{\hat W}_{ij}-
{1\over 2}\delta_{ij}{\hat W}_{kk}\right] +O(6) \ . \label{A.2c}
\end{eqnarray}
\end{mathletters}
Substituting the 1PN metric into the right-hand-side of the field equation we 
get

\begin{equation}
|g|= 1 + {4 \over c^2} V + {4\over c^4} \left({\hat W}_{kk}+2V^2\right)+O(6) 
\ , \label{A.3} 
\end{equation}
together with the gravitational source term (equations (2.12) in \cite{B95})

\begin{mathletters}
\label{A.4}
\begin{eqnarray}
 \Lambda^{00} &=& - {14\over c^4} \partial_i V
 \partial_i V + {16\over c^6} \biggl\{ - V \partial^2_t V - 2V_i \partial_t
 \partial_i V + {5\over 8} (\partial_t V)^2 \nonumber \\
&+& {1\over 2} \partial_i V_j (\partial_i V_j +3\partial_j V_i) 
 + \partial_i V\partial_t V_i - {7\over 2} V\partial_i V\partial_i V 
\nonumber\\ 
&-& \left({\hat W}_{ij}-{1\over 2}\delta_{ij}{\hat W}_{kk}\right) 
\partial^2_{ij} V - \partial_i V\partial_i {\hat 
W}_{kk}\biggr\} +O(8)  \ , \label{A.4a} \\
\Lambda^{0i} &=&  {16\over c^5} \left\{ \partial_j V
 (\partial_i V_j - \partial_j V_i) + {3\over 4} \partial_t V \partial_i V
 \right\} +O(7) \ , \label{A.4b} \\
 \Lambda^{ij} &=& {4\over c^4}\left\{ \partial_i V
 \partial_j V -{1\over 2} \delta_{ij} \partial_k V \partial_k V \right\}
\nonumber\\
  &+& {16\over c^6} \biggl\{ 2 \partial_{(i} V\partial_t V_{j)}
  - \partial_i V_k \partial_j V_k - \partial_k V_i \partial_k V_j
  + 2 \partial_{(i} V_k \partial_k V_{j)} \nonumber\\
&-& {3\over 8} \delta_{ij}
    (\partial_t V)^2 - \delta_{ij} \partial_k V \partial_tV_k 
  + {1\over 2} \delta_{ij}\partial_k V_m (\partial_k V_m
   - \partial_m V_k) \biggr\} +O(8) \ . \label{A.4c}
\end{eqnarray}
\end{mathletters}
These are the needed equations, which lead, by application of the retarded 
integral on the right-hand-side of the field equations, to 

\begin{mathletters}
\label{A.5}
\begin{eqnarray}
{h^{00} +h^{ii}\over 2}&=& -{2\over c^2} V -{4\over c^4} V^2
- {8\over c^6} \left[\hat{X}+{1\over 2} V {\hat W}_{ii}
+ {2V^3\over 3}\right] +O(8)  \ , \label{A.5a}\\
h^{0i} &=& -{4\over c^3} V_i - {8\over c^5} \left[ \hat{R}_i + V V_i\right]+ 
O(7) \ , \label{A.5b}\\
h^{ij} &=& -{4\over c^4} \left[{\hat W}_{ij}-
{1\over 2}\delta_{ij}{\hat W}_{kk}\right] +O(6)  \label{A.5c}
\end{eqnarray}
\end{mathletters}
(the potentials are defined in the text). From this, we deduce the components 
of the covariant metric $g_{\mu\nu}$ and find the result (\ref{2.5}). It has 
been shown in section III of \cite{B95} that the post-Newtonian metric 
matches in the external near zone to a solution extending up to the radiative 
zone.

\section{Complete results for the potentials}

We give first all the relevant potentials (valid all-over space-time) which 
are used in the obtention of the 2.5PN metric (\ref{7.2}):

\begin{mathletters}
\label{B.1}
\begin{eqnarray}
V &=& \frac{G m_1}{r_1} +\frac{G m_1}{c^2} \left(
-\frac{(n_1v_1)^2}{2r_1} +\frac{2v_1^2}{r_1} +
Gm_2 \left(-\frac{r_1}{4 r_{12}^3}-\frac{5}{4r_1r_{12}} +
\frac{r_2^2}{4r_1r_{12}^3}\right) \right) \nonumber \\
&+&\frac{2 G^2 m_1 m_2 (n_{12}v_{12})}{3c^3r_{12}^2}+
\frac{G m_1}{c^4r_1} \left( \frac{3 (n_1v_1)^4}{8}-
\frac{3(n_1v_1)^2 v_1^2}{2} + 2 v_1^4 \right) \nonumber \\
&+& \frac{G^2 m_1 m_2}{c^4} \left\{ v_1^2 \left( \frac{3 r_1^3}{16 r_{12}^5} - 
\frac{37 r_1}{16 r_{12}^3} - \frac{1}{r_1 r_{12}} - 
\frac{3 r_1 r_2^2}{16 r_{12}^5} + \frac{r_2^2}{r_1 r_{12}^3} \right) 
\right. \nonumber \\ & & \qquad \qquad +
v_2^2 \left( \frac{3 r_1^3}{16 r_{12}^5} + \frac{3 r_1}{16 r_{12}^3} + 
\frac{3}{2 r_1 r_{12}} - 
  \frac{3 r_1 r_2^2}{16 r_{12}^5} + \frac{r_2^2}{2 r_1 r_{12}^3} \right)
\nonumber \\ & & \qquad \qquad + 
(v_1v_2) \left(-\frac{3 r_1^3}{8 r_{12}^5} + \frac{13 r_1}{8 r_{12}^3} - 
\frac{3}{r_1 r_{12}} + 
  \frac{3 r_1 r_2^2}{8 r_{12}^5} - \frac{r_2^2}{r_1 r_{12}^3} \right)
\nonumber \\ & & \qquad \qquad
+(n_{12}v_1)^2 \left(-\frac{15 r_1^3}{16 r_{12}^5}+\frac{57 r_1}{16 r_{12}^3}+ 
  \frac{15 r_1 r_2^2}{16 r_{12}^5} \right)
\nonumber \\ & & \qquad \qquad
+(n_{12}v_2)^2 \left( -\frac{15 r_1^3}{16 r_{12}^5} - 
\frac{33 r_1}{16 r_{12}^3} + \frac{7}{8 r_1 r_{12}} + 
  \frac{15 r_1 r_2^2}{16 r_{12}^5} - \frac{3 r_2^2}{8 r_1 r_{12}^3} \right)
\nonumber \\ & & \qquad \qquad
+(n_{12}v_1)(n_{12}v_2) \left( \frac{15 r_1^3}{8 r_{12}^5} - 
\frac{9 r_1}{8 r_{12}^3} - \frac{15 r_1 r_2^2}{8 r_{12}^5} \right)
\nonumber \\ & & \qquad \qquad+
(n_1v_1)(n_{12}v_1) \left( -\frac{3 r_1^2}{2 r_{12}^4} + 
\frac{3}{4 r_{12}^2} + \frac{3 r_2^2}{4 r_{12}^4} \right)+
(n_{1}v_2)(n_{12}v_1) \left( \frac{3 r_1^2}{4 r_{12}^4}+
\frac{2}{r_{12}^2} \right) \nonumber \\ & & \qquad \qquad
+(n_1v_1)(n_{12}v_2) \left( \frac{3 r_1^2}{2 r_{12}^4} + 
\frac{13}{4 r_{12}^2} - \frac{3 r_2^2}{4 r_{12}^4} \right)+
(n_{1}v_2)(n_{12}v_2) \left( -\frac{3 r_1^2}{4 r_{12}^4} - 
\frac{3}{2 r_{12}^2} \right)
\nonumber \\ & & \qquad \qquad \left. +
(n_1v_1)^2 \left(-\frac{r_1}{8 r_{12}^3} +\frac{7}{8 r_1 r_{12}}
 - \frac{3 r_2^2}{8 r_1 r_{12}^3} \right)+
\frac{(n_1v_1) (n_1v_2) r_1}{2 r_{12}^3} \right\}
\nonumber \\ &+&
\frac{G^3 m_1^2 m_2}{c^4} \left( -\frac{r_1^3}{8 r_{12}^6} +
 \frac{5 r_1}{8 r_{12}^4} + \frac{3}{4 r_1 r_{12}^2} + 
  \frac{r_1 r_2^2}{8 r_{12}^6} - \frac{5 r_2^2}{4 r_1 r_{12}^4} \right)
\nonumber \\ &+&
\frac{G^3 m_1 m_2^2}{c^4} \left(-\frac{r_1^3}{32 r_{12}^6} + 
\frac{43 r_1}{16 r_{12}^4} + \frac{91}{32 r_1 r_{12}^2} - 
  \frac{r_1 r_2^2}{16 r_{12}^6} - \frac{23 r_2^2}{16 r_1 r_{12}^4} + 
  \frac{3 r_2^4}{32 r_1 r_{12}^6} \right)
\nonumber \\ &+&
\frac{G^2 m_1 m_2}{c^5} \left\{
(n_{12}v_{12})^3 \left( -\frac{7r_1^4}{4r_{12}^6} + 
\frac{3 r_1^2}{2 r_{12}^4} - \frac{11}{8 r_{12}^2} + 
 \frac{7 r_1^2 r_2^2}{4 r_{12}^6}\right) \right. \nonumber \\
& & \qquad \qquad
+(n_1v_{12}) (n_{12}v_{12})^2 \left( -\frac{2 r_1^3}{r_{12}^5} 
-\frac{5 r_1}{r_{12}^3} 
+ \frac{2 r_1 r_2^2}{r_{12}^5} \right)
-(n_1v_{12})^2 (n_{12}v_{12}) \frac{r_1^2}{r_{12}^4} 
\nonumber \\ & & \qquad \qquad 
+(n_{12}v_1)(n_{12}v_{12})^2 \left(\frac{5 r_1^2}{r_{12}^4} 
+ \frac{5}{r_{12}^2} -\frac{5 r_2^2}{r_{12}^4} \right)
+(n_1v_{12})(n_{12}v_1)(n_{12}v_{12})\frac{6r_1}{r_{12}^3}
\nonumber \\ & & \qquad \qquad
+(n_1v_1)(n_{12}v_{12})^2 \frac{2r_1}{r_{12}^3}
-\frac{(n_{12}v_1)^2 (n_{12}v_{12})}{r_{12}^2}
-(n_1v_{12})v_1^2 \frac{8r_1}{3r_{12}^3}
+(n_{12}v_1) v_1^2 \frac{8}{3r_{12}^2}
\nonumber \\ & & \qquad \qquad
+ (n_{12}v_{12}) v_1^2 \left(-\frac{3 r_1^2}{r_{12}^4}
-\frac{8}{3 r_{12}^2} + \frac{3 r_2^2}{r_{12}^4} \right)
+(n_{12}v_{12})(v_1v_2) \frac{7}{3 r_{12}^2} 
\nonumber \\ & & \qquad \qquad
+(n_1v_{12})(v_1v_2)\frac{8r_1}{3r_{12}^3}
-(n_{12}v_{1})(v_1v_2)\frac{8}{3r_{12}^2}
-(n_1v_1)v_{12}^2 \frac{2r_1}{3r_{12}^3}
\nonumber \\ & & \qquad \qquad
+(n_{12}v_{12})v_{12}^2 \left( \frac{3 r_1^4}{4 r_{12}^6} - 
\frac{9 r_1^2}{10 r_{12}^4} - \frac{9}{8 r_{12}^2} - 
  \frac{3 r_1^2 r_2^2}{4 r_{12}^6} \right)
\nonumber \\ & & \qquad \qquad \left.
+(n_1v_{12})v_{12}^2 \left(\frac{2 r_1^3}{5 r_{12}^5}+\frac{5 r_1}{3 r_{12}^3} 
- \frac{2 r_1 r_2^2}{5 r_{12}^5} \right)
+(n_{12}v_1)v_{12}^2 \left( -\frac{r_1^2}{r_{12}^4} - \frac{5}{3 r_{12}^2} 
+ \frac{r_2^2}{r_{12}^4} \right) \right\}
\nonumber \\ &+&\frac{G^3 m_1^2 m_2}{c^5} \left\{
(n_1v_{12}) \left( -\frac{4 r_1^3}{15 r_{12}^6}- \frac{8 r_1}{3r_{12}^4} + 
  \frac{4 r_1 r_2^2}{15 r_{12}^6} \right)
+(n_{12}v_1) \left( -\frac{2 r_1^2}{3 r_{12}^5} + \frac{10}{3 r_{12}^3} + 
\frac{2 r_2^2}{3 r_{12}^5} \right) \right.
\nonumber \\ & & \qquad \qquad \left.
+(n_1v_1) \frac{4 r_1}{3r_{12}^4}
+(n_{12}v_{12}) \left(-\frac{2r_1^2}{3 r_{12}^5} - \frac{20}{3 r_{12}^3} 
+ \frac{8}{3 r_1 r_{12}^2} + \frac{34 r_2^2}{15 r_{12}^5} \right) \right\}
\nonumber \\ &+& O(6)+1\leftrightarrow 2  \ , \label{B.1a} \\
V_i &=&  \frac{G m_1 v_1^i}{r_1}+
       n^i_{12} \frac{G^2 m_1 m_2}{c^2 r_{12}^2} \left((n_1v_1)+ 
\frac{3 (n_{12}v_{12}) r_1}{2 r_{12}}\right)
\nonumber \\ &+& \frac{v_1^i}{c^2} \left\{
\frac{G m_1}{ r_1} \left(-\frac{(n_1v_1)^2}{2} + v_1^2 \right)
+ G^2 m_1 m_2 \left( -\frac{3 r_1}{4 r_{12}^3} + 
\frac{r_2^2}{4 r_1 r_{12}^3} - \frac{5}{4 r_1 r_{12}} \right) \right\}
+v_2^i \frac{G^2 m_1 m_2 r_1}{2 c^2 r_{12}^3} \nonumber \\ 
&+& \frac{n_{12}^i}{c^3} G^2 m_1 m_2 \left[
(n_{12}v_{12})^2 \frac{5 r_1^2}{2 r_{12}^4}
+(n_{12}v_{12})(n_1v_{12}) \frac{3 r_1}{2 r_{12}^3} 
-v_{12}^2 \frac{r_1^2}{2 r_{12}^4} +\frac{v_1^2}{2 r_{12}^2} \right.
\nonumber \\ & & \left. \qquad 
+ G m_1 \left( -\frac{r_1^2}{6 r_{12}^5} + \frac{r_2^2}{6 r_{12}^5} + 
\frac{1}{2 r_{12}^3} \right) \right]
-\frac{G^2 m_1 m_2 v_1^i}{3c^3 r_{12}^2} \nonumber \\
&+&\frac{G^2 m_1 m_2 v_{12}^i}{c^3 r_{12}^2} \left(
- \frac{3 r_1^2}{2 r_{12}^4}(n_{12}v_{12}) -
\frac{2 r_1}{3 r_{12}^3} (n_1v_{12}) +\frac{(n_{12}v_1)}{2 r_{12}^2} \right)
+ O(4)+1\leftrightarrow 2 \ , \label{B.1b} \\
{\hat W}_{ij} &=&
\delta^{ij} \left(  -\frac{G m_1 v_1^2}{r_1}-\frac{G^2 m_1^2}{4 r_1^2}+
\frac{G^2 m_1 m_2}{r_{12} S} \right)
+\frac{G m_1 v_1^i v_1^j}{r_1} +\frac{G^2 m_1^2 n_1^i n_1^j}{4 r_1^2}
\nonumber \\ &+&
G^2 m_1 m_2 \left\{ \frac{1}{S^2} \left(n_1^{(i} n_2^{j)} 
+ 2n_1^{(i} n_{12}^{j)} \right)-
n_{12}^i n_{12}^j \left(\frac{1}{S^2}+\frac{1}{r_{12} S} \right) \right\}
\nonumber \\ &+&
\frac{G^2 m_1 m_2}{c r_{12}^2} \left(-\frac{(n_{12}v_{12})}{2} \delta^{ij}-
\frac{3 (n_{12}v_{12})}{2} n_{12}^i n_{12}^j + 
2n_{12}^{(i} v_{12}^{j)} \right)
\nonumber \\ &+&O(2)+1\leftrightarrow 2 \ , \label{B.1c} \\
{\hat R}_i &=&
G^2 m_1 m_2 n_{12}^i \left\{ -\frac{(n_{12}v_1)}{2S} \left(\frac{1}{S}+
\frac{1}{r_{12}} \right)
-\frac{2 (n_2v_1)}{S^2}+\frac{3(n_2v_2)}{2S^2} \right\}
\nonumber \\ &+&
n_1^i \left\{ \frac{G^2 m_1^2 (n_1v_1)}{8 r_1^2}
+\frac{G^2 m_1 m_2}{S^2} \left( 2 (n_{12}v_1) - \frac{3(n_{12}v_2)}{2}
+2 (n_2v_1) - \frac{3(n_2v_2)}{2} \right) \right\}
\nonumber \\ &+& 
v_1^i \left\{ -\frac{G^2 m_1^2}{8r_1^2}+
G^2 m_1 m_2 \left( \frac{1}{r_1 r_{12}} + \frac{1}{2 r_{12}S} \right) \right\}
-v_2^i\frac{G^2 m_1 m_2}{r_1 r_{12}} 
+ v_1^i \frac{G^2 m_1 m_2 (n_{12}v_1)}{2c r_{12}^2}
\nonumber \\ &+& n_{12}^i \frac{G^2 m_1 m_2}{c r_{12}^2} \left(
-\frac{3(n_{12}v_1)^2}{4} + \frac{v_1^2}{4} \right)
+ O(2)+1\leftrightarrow 2 \ , \label{B.1d} \\
{\hat X} &=&
\frac{G^2 m_1^2}{8r_1^2} \left((n_1v_1)^2-v_1^2\right)
+G^2 m_1 m_2 v_1^2 \left(\frac{1}{r_1 r_{12}} 
+ \frac{1}{r_1 S} + \frac{1}{r_{12} S} \right)
\nonumber \\ &+&
G^2 m_1 m_2 \left\{v_2^2 \left(-\frac{1}{r_1 r_{12}} + \frac{1}{r_1 S} 
+ \frac{1}{r_{12} S} \right)
-\frac{(v_1v_2)}{S} \left(\frac{2}{r_1}+
\frac{3}{2 r_{12}} \right)
-\frac{(n_{12}v_1)^2}{S} \left(\frac{1}{S}+\frac{1}{r_{12}}\right) \right.
\nonumber \\ & & \qquad \qquad
-\frac{(n_{12}v_2)^2}{S} \left(\frac{1}{S} + \frac{1}{r_{12}} \right)
+\frac{3(n_{12}v_1)(n_{12}v_2)}{2S} \left(\frac{1}{S} + 
\frac{1}{r_{12}}\right)+\frac{2(n_{12}v_1)(n_1v_1)}{S^2}
\nonumber \\ & & \qquad \qquad
-\frac{5(n_{12}v_2)(n_1v_1)}{S^2}
-\frac{(n_1v_1)^2}{S} \left(\frac{1}{S} + \frac{1}{r_1} \right)
+\frac{2(n_{12}v_2)(n_1v_2)}{S^2} \nonumber \\ & & \qquad \qquad
+\frac{2(n_1v_1)(n_1v_2)}{S} \left(\frac{1}{S} + \frac{1}{r_1} \right)
-\frac{(n_1v_2)^2}{S} \left(\frac{1}{S} + \frac{1}{r_1}\right)
-\frac{2(n_{12}v_2)(n_2v_1)}{S^2} \nonumber \\ 
& & \left. \qquad \qquad +\frac{2(n_1v_2)(n_2v_1)}{S^2}
-\frac{3(n_1v_1)(n_2v_2)}{2 S^2} \right\}+\frac{G^3 m_1^3}{12 r_1^3}
\nonumber \\ 
&+& G^3 m_1^2 m_2 \left(\frac{1}{2 r_1^3} + \frac{1}{16 r_2^3} + 
\frac{1}{16 r_1^2 r_2} - \frac{r_2^2}{2 r_1^2 r_{12}^3} + 
\frac{r_2^3}{2 r_1^3 r_{12}^3} - \frac{r_1^2}{32 r_2^3 r_{12}^2} 
- \frac{3}{16 r_2 r_{12}^2} + 
\frac{15 r_2}{32 r_1^2 r_{12}^2} \right. \nonumber \\ 
& & \left. \qquad \qquad - \frac{r_2^2}{2 r_1^3 r_{12}^2} 
- \frac{r_2}{2 r_1^3 r_{12}} - \frac{r_{12}^2}{32 r_1^2 r_2^3} \right)
+G^3 m_1 m_2^2 \left( -\frac{1}{2 r_{12}^3} + \frac{r_2}{2 r_1 r_{12}^3} 
-\frac{1}{2 r_1 r_{12}^2} \right) \nonumber \\ 
&+& \frac{G^2 m_1 m_2}{cr_{12}^2} \left( -\frac{3(n_{12}v_1)^2 
(n_{12}v_{12})}{4}+ 
  \frac{3 (n_{12}v_1) (n_{12}v_{12})^2}{4} -  
  \frac{3 (n_{12}v_{12})^3}{2} - \frac{(n_{12}v_1) v_{12}^2}{4}
\right. \nonumber \\ & &\qquad \qquad  \left.
+ \frac{3 (n_{12}v_{12}) v_{12}^2}{2} + 
  \frac{(n_{12}v_1) v_1^2}{2} - \frac{(n_{12}v_{12}) v_1^2}{4} 
- \frac{(n_{12}v_1) (v_1v_2)}{2} + 
  \frac{(n_{12}v_{12}) (v_1v_2)}{2} \right) \nonumber \\ 
&+& \frac{G^3 m_1^2 m_2}{c} \left\{
(n_{12}v_{12}) \left( \frac{3}{8 r_1^3} + \frac{3 r_1}{8 r_{12}^4} 
-\frac{3 r_2^2}{4 r_1 r_{12}^4} + 
\frac{3 r_2^4}{8 r_1^3 r_{12}^4} - \frac{3}{2 r_{12}^3} 
+ \frac{1}{4 r_1 r_{12}^2} - 
  \frac{3 r_2^2}{4 r_1^3 r_{12}^2}\right) \right.
\nonumber \\ & & \left. \qquad \qquad
+(n_1v_{12}) \left( \frac{1}{r_{12}^3} - \frac{r_2^2}{r_1^2 r_{12}^3} 
+ \frac{1}{r_1^2 r_{12}} \right) \right\}
+O(2)+1\leftrightarrow 2 \ . \label{B.1e} 
\end{eqnarray}
\end{mathletters}
The values of the potentials at the location of body 1, following from 
(\ref{3.6}), are

\begin{mathletters}
\label{B.2} 
\begin{eqnarray}
(V)_1 &=& {Gm_2\over r_{12}} \left\{1+{1\over c^2}\left[-{3\over 2}
{Gm_1\over r_{12}}+2v_2^2-{1\over 2}(n_{12}v_2)^2\right]
+{4\over 3}{Gm_1\over r_{12}c^3}(n_{12}v_{12}) \right.\nonumber\\
& &\qquad  +{Gm_1\over r_{12}c^4}\left[ {11\over 2}{Gm_1\over r_{12}}+
{5\over 4}{Gm_2\over r_{12}}+{15\over 
8}v_1^2-{7\over 4}(v_1v_2)-{25\over 8}v_2^2 \right. \nonumber\\
& &\qquad \qquad \quad \left.+ {1\over 8}(n_{12}v_1)^2-
{25\over 4}(n_{12}v_1)(n_{12}v_2)+{33\over 8}(n_{12}v_2)^2\right] \nonumber\\
& &\left. \qquad+{1\over c^4}\left[2v_2^4-{3\over 2}(n_{12}v_2)^2v_2^2+
{3\over 8}(n_{12}v_2)^4\right]\right\}\nonumber\\
&+&{G^2m_1m_2\over c^5r_{12}^2}\left\{ -{2\over 5}{Gm_1\over r_{12}}
(n_{12}v_1) -{26\over 15}{Gm_2\over r_{12}}(n_{12}v_1)
+{22\over 5}{Gm_1\over r_{12}}(n_{12}v_2)\right.\nonumber\\
& &\qquad \qquad+{2\over 5}{Gm_2\over r_{12}}(n_{12}v_2)
-{32\over 15}(n_{12}v_1)v_1^2
+{48\over 5}(n_{12}v_1)(v_1v_2)-{122\over 15}(n_{12}v_1)v_2^2\nonumber\\
& &\qquad \qquad+{92\over 15}(n_{12}v_2)v_1^2-{184\over 15}
(n_{12}v_2)(v_1v_2)+{34\over 5}(n_{12}v_2)v_2^2
+2(n_{12}v_1)^3\nonumber\\
& &\qquad \qquad \left.- 10(n_{12}v_1)^2(n_{12}v_2)
+12(n_{12}v_1)(n_{12}v_2)^2
-4(n_{12}v_2)^3\right\}+O(6)   \ , \label{B.2a} 
\\
(V_i)_1 &=& {Gm_2\over r_{12}}\biggl\{v_2^i
+{v_2^i\over c^2}\left[-2{Gm_1\over r_{12}}+v_2^2-
{1\over 2}(n_{12}v_2)^2\right] +{1\over 2}{Gm_1\over 
r_{12}c^2}v_1^i \nonumber\\
& &\qquad+{Gm_1\over r_{12}c^2}n_{12}^i\left[-{3\over 2}(n_{12}v_1)+
{1\over 2}(n_{12}v_2)\right]\nonumber\\
& &\qquad+{Gm_1\over c^3r_{12}}v_1^i(n_{12}v_1)
-{Gm_1\over c^3r_{12}}v_2^i\left[{5\over 3}(n_{12}v_1)-
{2\over 3}(n_{12}v_2)\right]\nonumber\\
& &\qquad+{Gm_1\over c^3r_{12}}n_{12}^i\left[{2\over 3}{Gm_1\over r_{12}}
-{1\over 3}{Gm_2\over r_{12}}+v_1^2-(v_1v_2)-(n_{12}v_{12})^2 \right]
\biggr\}+O(4) \ , \label{B.2b} 
\\
({\hat W}_{ij})_1 &=& {Gm_2\over r_{12}}\biggl\{v_2^{ij}-\delta^{ij}v_2^2+
{Gm_1\over r_{12}}\left[-2n_{12}^{ij}+\delta^{ij}\right]\nonumber\\
&&\qquad+{Gm_2\over 4r_{12}}\left[n_{12}^{ij}-\delta^{ij}\right]+4{Gm_1\over 
cr_{12}}n_{12}^{(i}v_{12}^{j)}\nonumber\\
&&\qquad-{Gm_1\over cr_{12}}(n_{12}v_{12})\left(3n_{12}^{ij}+\delta^{ij}\right)
\biggr\}+O(2) \ , \label{B.2c} \\
({\hat R}_i)_1 &=& {G^2m_1m_2\over r_{12}^2}\left[-{3\over 4}v_1^i+
{5\over 4}v_2^i-{1\over 
2}(n_{12}v_1)n_{12}^i-{1\over 2}(n_{12}v_2)n_{12}^i\right] \nonumber\\
&+&{G^2m_2^2\over r_{12}^2}\left[-{1\over 8}v_2^i+
{1\over 8}(n_{12}v_2)n_{12}^i\right] \nonumber\\
&+&{G^2m_1m_2\over cr_{12}^2}\biggl\{n_{12}^i\left[ {1\over 4}v_1^2-
{1\over 4}v_2^2-{3\over 4}(n_{12}v_1)^2+{3\over 
4}(n_{12}v_2)^2\right]\nonumber\\
&&\qquad +{1\over 2}(n_{12}v_1)v_1^i-{1\over 2}(n_{12}v_2)v_2^i\biggr\}+O(2)
\ , \label{B.2d} \\
({\hat X})_1 &=& {G^2m_1m_2\over r_{12}^2}\left[-{3\over 2}{Gm_1\over r_{12}}+
{1\over 4}v_1^2-
2(v_1v_2)+{9\over 4}v_2^2 \right.\nonumber\\
&&\quad-\left.{11\over 4}(n_{12}v_1)^2+{9\over 2}(n_{12}v_1)(n_{12}v_2)
-{11\over 4}(n_{12}v_2)^2\right] \nonumber\\
&+&{G^2m_2^2\over r_{12}^2}\left[{1\over 12}{Gm_2\over r_{12}}-
{1\over 8}v_2^2+{1\over 8}(n_{12}v_2)^2\right]\nonumber\\
&+&\frac{G^2 m_1 m_2}{cr_{12}^2} \biggl\{ -3 (n_{12}v_1)^3 + 
\frac{15}{2}(n_{12}v_1)^2 (n_{12}v_2) -
\frac{15}{2} (n_{12}v_1) (n_{12}v_2)^2 + 3 (n_{12}v_2)^3 \nonumber\\
&&\quad+ 3 (n_{12}v_1) v_1^2 - \frac{5}{2} (n_{12}v_2) v_1^2  
- 5 (n_{12}v_{12}) (v_1v_2) + \frac{5}{2} (n_{12}v_1) v_2^2 - 
3 (n_{12}v_2) v_2^2 \nonumber \\ 
&& \quad-\frac{3}{2} \frac{G m_1}{r_{12}} (n_{12}v_{12}) 
-\frac{5}{2} \frac{G m_2}{ r_{12}}(n_{12}v_{12}) \bigg\}
+O(2) \ . \label{B.2e} 
\end{eqnarray}
\end{mathletters}
The gradients of the potentials computed at body 1 (needed for the equations 
of motion) are

\begin{mathletters}
\label{B.3}
\begin{eqnarray}
(\partial_iV)_1 &=& - {Gm_2\over r_{12}^2} n_{12}^i
 +{Gm_2\over r_{12}^2c^2} n_{12}^i \left[ - 2v^2_2 
  + {3\over 2} (n_{12}v_2)^2 + {Gm_1\over r_{12}}\right]-
{Gm_2\over r_{12}^2c^2}(n_{12}v_2)v_2^i \nonumber\\
&+&{Gm_2\over r_{12}^2c^4} n_{12}^i \left[ -2v^4_2 + 
{9\over 2} v^2_2 (n_{12}v_2)^2 -{15\over 8}(n_{12}v_2)^4 \right] \nonumber\\ 
&+&{G^2m_1m_2\over r_{12}^3c^4}n_{12}^i \left[ -{3\over 4} v^2_1 -
{3\over 4} v^2_2+{7\over 2} (v_1v_2) - {13\over 2} (n_{12}v_1)^2 \right.
\nonumber\\
&&\qquad \qquad \quad \left. +7(n_{12}v_1)(n_{12}v_2)+
{1\over 2}(n_{12}v_2)^2 -{1\over 4}{Gm_1\over r_{12}}-
{1\over 2}{Gm_2\over r_{12}} \right]\nonumber\\
 &+& {Gm_2\over r_{12}^2c^4}v_2^i\left[ -3v^2_2(n_{12}v_2)+
{3\over 2}(n_{12}v_2)^3 -{17\over 4}{Gm_1\over 
r_{12}}(n_{12}v_1)+{9\over 4}{Gm_1\over r_{12}}(n_{12}v_2)\right]\nonumber\\
&+& {G^2m_1m_2\over r_{12}^3c^4} v_1^i\left[ -{9\over 4}(n_{12}v_2) + 
{9\over 4}(n_{12}v_1)\right] \nonumber \\ 
&+& \frac{G^2 m_1 m_2}{c^5 r_{12}^3} n_{12}^i \left(-10 (n_{12}v_1)^3 + 
  10 (n_{12}v_1)^2 (n_{12}v_2) + 10 (n_{12}v_1) (n_{12}v_2)^2 \right. 
\nonumber \\ & & \qquad \qquad \quad  - 
  10 (n_{12}v_2)^3 +\frac{42}{5} (n_{12}v_1) v_1^2
- \frac{22}{5} (n_{12}v_2) v_1^2 - \frac{24}{5} (n_{12}v_1) (v_1v_2) 
\nonumber \\ & & \qquad \qquad \quad  - 
 \frac{16}{5} (n_{12}v_2) (v_1v_2) - 
  \frac{18}{5} (n_{12}v_1) v_2^2 + \frac{38}{5} (n_{12}v_2) v_2^2
+ \frac{88}{15} \frac{G m_1}{r_{12}} (n_{12}v_1) \nonumber \\ 
& & \qquad \qquad \quad \left.
- \frac{52}{15}\frac{G m_2}{r_{12}}(n_{12}v_1)  - 
  \frac{68}{15} \frac{G m_1}{r_{12}} (n_{12}v_2)+ 
  \frac{24}{5} \frac{G m_2}{r_{12}}  (n_{12}v_2)\right) \nonumber \\ 
&+& \frac{G^2 m_1 m_2}{c^5 r_{12}^3} v_1^i 
\left(10 (n_{12}v_1)^2 - 8 (n_{12}v_1) (n_{12}v_2) - 2 (n_{12}v_2)^2 - 
  \frac{16}{15} \frac{G m_1}{r_{12}} + 
  \frac{44}{15} \frac{G m_2}{r_{12}} \right. 
\nonumber \\ & & \qquad \qquad \quad \left. - \frac{62}{15} v_1^2 + 
\frac{44}{15} (v_1v_2) + \frac{6}{5} v_2^2 \right)
\nonumber \\ 
&+&\frac{G^2 m_1 m_2}{c^5 r_{12}^3} v_2^i \left(-6 (n_{12}v_1)^2 + 
6 (n_{12}v_2)^2 + \frac{12}{5} \frac{G m_1}{r_{12}} - 
\frac{8}{5} \frac{G m_2}{r_{12}} + \frac{14}{5} v_1^2 \right.
\nonumber \\ & & \qquad \qquad \quad \left. - \frac{4}{15} (v_1v_2)-
 \frac{38}{15} v_2^2\right) +O(6) \ , \label{B.3a}\\
(\partial_jV_i)_1 &=& -{Gm_2\over r_{12}^2}n_{12}^jv_2^i+
{Gm_2\over r_{12}^2c^2}\left\{n_{12}^{ij}{Gm_1\over r_{12}}\left[ -{3\over 2}
(n_{12}v_1) + {5\over 2}(n_{12}v_2)\right]\right.\nonumber\\
& & \qquad 
+{Gm_1\over 2r_{12}}n_{12}^jv_1^i+n_{12}^jv_2^i\left[-v_2^2+{3\over 2}
(n_{12}v_2)^2+{Gm_1\over 2r_{12}}\right]\nonumber\\
& & \qquad \left.-{Gm_1\over r_{12}}n_{12}^iv_2^j-v_2^{ij}(n_{12}v_2)\right\}
\nonumber \\
&+& \frac{G^2 m_1 m_2}{c^3 r_{12}^3} \biggl\{ n_{12}^{ij} \left( 
-5 (n_{12}v_{12})^2 + v_{12}^2
+ \frac{1}{3}\frac{G m_1}{r_{12}} + 
\frac{1}{3} \frac{G m_2}{r_{12}} \right)\nonumber\\
&&\qquad +6 n_{12}^{(i} v_1^{j)} (n_{12}v_{12})
-6 n_{12}^{(i} v_2^{j)} (n_{12}v_{12})
-\frac{4}{3} v_{12}^{ij}\biggr\}+O(4) \ , \label{B.3b} \\
(\partial_k{\hat W}_{ij})_1 &=& {Gm_2\over r_{12}^2}\biggl\{-n_{12}^kv_2^{ij}+
\delta^{ij}n_{12}^k\left[-{Gm_1\over 2r_{12}}+{Gm_2\over 2r_{12}}+v_2^2\right] 
\nonumber\\
&&\quad\quad+\delta^{k(i}n_{12}^{j)}\left[-{3Gm_1\over 2r_{12}}+{Gm_2\over 2r_{
12}}\right] 
+n_{12}^{ijk}\left[{2Gm_1\over r_{12}}-{Gm_2\over r_{12}}\right]\biggr\} +O(2) 
\ , \label{B.3c}\\
(\partial_j{\hat R}_i)_1 &=& 
{Gm_2\over r_{12}^2}\biggl\{\delta^{ij}\left[-{5\over 8}{Gm_1\over r_{12}}(n_{1
2}v_1)+{1\over 4}{Gm_1\over r_{12}}(n_{12}v_2)+
{1\over 8}{Gm_2\over r_{12}}(n_{12}v_2)\right]\nonumber\\
&&\quad\quad +n_{12}^{ij}\left[{1\over 2}{Gm_1\over r_{12}}(n_{12}v_1)+
{1\over 2}{Gm_1\over r_{12}}(n_{12}v_2)-
{1\over 2}{Gm_2\over r_{12}}(n_{12}v_2)\right]\nonumber\\
&&\quad\quad +n_{12}^i\left[{1\over 4}{Gm_1\over r_{12}}v_1^j-
{5\over 8}{Gm_1\over r_{12}}v_2^j+
{1\over 8}{Gm_2\over r_{12}}v_2^j\right]\nonumber\\
&&\quad\quad +n_{12}^j\left[{7\over 8}{Gm_1\over r_{12}}v_1^i-
{9\over 8}{Gm_1\over r_{12}}v_2^i+
{1\over 4}{Gm_2\over r_{12}}v_2^i\right]\biggr\}+O(2) \ , \label{B.3d}
\\
(\partial_i{\hat X})_1 &=& 
{G^2m_1m_2\over r_{12}^3}\biggl\{n_{12}^{i}\left[{1\over 2}{Gm_1\over r_{12}}-
{3\over 2}{Gm_2\over r_{12}}+
{9\over 2}(n_{12}v_1)^2-8(n_{12}v_1)(n_{12}v_2)\right.\nonumber\\
& &\qquad \qquad \quad +\left.{9\over 2}(n_{12}v_2)^2
+{7\over 4}(v_1v_2)-2v_2^2\right]\nonumber\\
&&\quad\quad -{5\over 2}(n_{12}v_1)v_1^i
+3(n_{12}v_2)v_1^i+{5\over 4}(n_{12}v_1)v_2^i-
{5\over 2}(n_{12}v_2)v_2^i\biggr\}\nonumber\\
&+&{G^2m_2^2\over r_{12}^3}\biggl\{n_{12}^{i}\left[-{1\over 4}
{Gm_2\over r_{12}}- {1\over 2}(n_{12}v_2)^2+{1\over 4}v_2^2\right]
+{1\over 4}(n_{12}v_2)v_2^i\biggr\}\nonumber\\
&+&\frac{G^3 m_1 m_2^2}{cr_{12}^4}\biggl\{ 3n_{12}^i (n_{12}v_{12})
- 2v_{12}^i\biggr\} +O(2) \ . \label{B.3e}
\end{eqnarray}
\end{mathletters}


\begin{references}
\bibitem{N1}Some poles are expected to develop at the next 3PN approximation, 
corresponding to the appearance of logarithmic terms in the equations of 
motion.
\bibitem{N2}The post-Newtonian metric though valid formally everywhere in 
space-time is expected to constitute a good approximation to an exact 
solution only in the near zone of the source, i.e. in a region of much 
smaller scale than the wavelength of the emitted radiation. For inspiralling 
compact binaries, the near-zone covers entirely the binary's orbit.
\bibitem{N3}Greek indices take the values 0,1,2,3, and Latin 1,2,3. Our 
signature is (--+++).
\bibitem{Note}Indeed, using the equation of continuity (\ref{2.2a}), we see 
that there is no term $O(1)$.
\bibitem{N4}We have ${\hat W}_{ij}=W_{ij}-\delta_{ij}W_{kk}$, where $W_{ij}$ 
is the definition adopted in \cite{B95}.
\bibitem{N5}Since all infinite terms are removed the Hadamard regularization 
does not need to be followed by a renormalization.
\bibitem{N6}A less trivial example (not needed in this paper) is 
$(V^4)_1=G^4m_2^2(2m_1^2+m_2^2)/r_{12}^4+O(2)$, showing that the Hadamard 
partie finie is in general not ``distributive'' with respect to 
multiplication, i.e. $(V^4)_1$ is not equal to $[(V)_1]^4$.
\bibitem{N12}Note that many of the ``odd'' terms in the metric, which are 
associated with radiation-reaction, are only functions of time $t$ or depend 
smoothly on the field point ${\bf x}$ (through for instance the product 
$r_1{\bf n}_1={\bf x}-{\bf y}_1$). However, when presented into a form which 
shows this fact more explicitly (expressing all the terms with the help of 
the vectors ${\bf r}_1$ and ${\bf r}_{12}$) the metric is not simpler.
\bibitem{N13}We have also 
$(nv_{12})=r_{12}~\omega_{\rm 2PN}\sin i\cos \phi_{\rm 2PN}+O(5)$, 
where $\phi_{\rm 2PN}$ denotes the orbital phase and $i$ the inclination 
angle.
\end{references}
\end{document}